\documentclass[onecolumn,authoryear]{els-mrw} 

\usepackage{amsmath,amssymb,amsfonts,amsthm,makeidx,graphicx}
\usepackage{txfonts}
\usepackage{helvet}
\usepackage[utf8]{inputenc}
\usepackage{comment}
\usepackage{multirow}
\usepackage{booktabs,array,amssymb,makecell,graphicx,xcolor,longtable,ragged2e,pdflscape}
\newcommand{\yes}{\checkmark}
\newcommand{\no}{--}
\newcommand{\mc}{MC}

\newcommand{\fixed}{Fixed}
\newcommand{\nn}{NN}
\newcolumntype{C}[1]{>{\centering\arraybackslash}p{#1}}
\newcolumntype{L}[1]{>{\raggedright\arraybackslash}p{#1}}

\usepackage[colorlinks,citecolor=blue]{hyperref}

\begin{document}
\chapter{Hadronization in QCD and Fragmentation Functions}
\author[1]{Ignacio Borsa}%
\author[1]{Marco Stratmann}%
\address[1]{\orgname{Institute for Theoretical Physics}, 
\orgdiv{University of T\"ubingen}, 
\orgaddress{Auf der Morgenstelle 14, 72076 T\"ubingen, Germany}}
%
\maketitle
%




\begin{abstract}[Abstract]
The hadronization of quarks and gluons is reviewed within the framework of collinear
parton-to-hadron fragmentation functions closely associated
with the notion of factorization in hard scattering processes in QCD.
The state-of-the-art in calculating both the
scale evolution of fragmentation functions
and the partonic cross sections pertinent to their determination from data
up to next-to-next-leading order accuracy is presented. Potential
improvements up to next-to-next-to-leading logarithmic order are outlined as well.
All relevant experimental data sets for single-inclusive
pion, kaon, and unidentified charged hadron production are listed.
The most recent sets of fragmentation functions and their uncertainties
extracted from these data in global QCD analyses utilizing various different strategies are presented and compared to each other.
\end{abstract}
\section{Introduction: Why and where do we need fragmentation functions 
and how do we define them?}\label{chap:intro}
Quantum Chromodynamics (QCD) is the well-established theory describing the 
strong interactions of color-charged (anti)quarks and gluons, 
the fundamental building blocks of matter. 
Yet, in nature, i.e., in our particle detectors, due to confinement
only colorless bound-states, mesons and baryons, have been observed so far. 
Theoretical estimates of relevant cross sections or hadron yields thus require
the modeling of the hadronization of quarks and gluons.
That is where fragmentation functions come into play.

More specifically, one invokes the framework of perturbative QCD (pQCD) and the idea
that a cross section calculation comprises several fundamental building
blocks, some of them calculable as a series expansion in the strong coupling,
$\alpha_S(\mu)\ll 1$, while others are genuinely non-perturbative but universal and
hence quantifiable from data, a crucial property.
The foundations of this approach have been properly established 
by nontrivial factorization theorems  which make the separation of 
long and short distance/time scales in QCD scattering processes manifest 
if the latter are characterized by at least one large energy scale $\mu$ of the 
order of a few GeV \cite{Collins:1989gx,Collins:2011zzd}.

Applying the concept of factorization to single-inclusive 
electron-positron annihilation (SIA), $e^+e^-\to \gamma^*/Z\to hX$, 
into a detected hadron $h$ and some experimentally unidentified 
hadronic remnant $X$ inevitably leads to the introduction 
of a set of non-perturbative parton-to-hadron fragmentation functions (FFs)
$D_i^h(z,\mu)$ for each (anti)quark flavor $i=q,\bar{q}$ and the gluon $i=g$
\cite{Collins:1981uk,Collins:1981uw,Collins:2011zzd}.
In fact, it turns out that the SIA cross section can be written 
as a sum of convolutions of the $D_i^h(z,\mu)$ with calculable coefficient functions 
encoding the partonic interactions at short distances, $\gamma^*\to q\bar{q}$, 
$\gamma^*\to q\bar{q}g$, etc.
The necessary hard scale $\mu$ for SIA is provided by the 
center-of-mass system (c.m.s.) energy $\sqrt{S}$ of the $e^+e^-$ collisions
which determines also the virtuality of the exchanged intermediate 
virtual photon $\gamma^*$ or $Z$ boson.

This is all reminiscent of the steps taken to analyze the space-like counterpart
of SIA, deep-inelastic lepton-nucleon scattering (DIS), $l N\to \gamma^*/Z \to X$,
in terms of calculable partonic coefficient functions and a set of non-perturbative 
parton distribution functions (PDFs) $f^h_i(x,\mu)$.
The physical interpretation of FFs $D_i^h(z,\mu)$ also closely
resembles that for PDFs $f^h_i(x,\mu)$. 
While the latter determine the distribution of partons $i$ 
with fractional momentum $x$ inside a hadron $h$ that enters  
a high-energy collision, FFs can be viewed, in the naive parton model, as the probability density 
that a massless (anti)quark or gluon produced in a hard-scattering
forms a hadron $h$ that takes a fraction $z$ of its momentum.
In case of SIA, $z$ simply denotes the scaled energy $z=2E_h/\sqrt{S}$
of the hadron. 

In this picture, the hadronization process is assumed to be independent of any 
other colored particles produced in the hard scattering, and it is not
specified nor relevant how the color neutralization of a fragmenting 
colorful parton actually takes place.
If one wants to model the entire hadronic final-state of a high-energy collision
rather than to estimate the inclusive momentum or energy distribution of one or 
- at most - two observed hadrons, for which the concept of FFs applies, 
one has to resort to one of the several available Monte Carlo event generators
\cite{Sjostrand:2014zea,Bewick:2023tfi,Sherpa:2024mfk}.
They all have in common that after the calculable hard scattering
an embedded parton shower generates ensembles of partons
of, in principle, arbitrarily high multiplicity.
To hadronize them either cluster or string fragmentation
is invoked, both tuned to experimental data by adjusting 
a large number of free parameters.
While the former thrives on the non-perturbative splitting of the
mainly soft gluons at the end of the shower into $q\bar{q}$ pairs
and the subsequent decay of these clusters into
hadrons, the main idea of the latter model is to produce the hadrons in
a sequence of multiple string breakings, where the string or 
flux tube resembles the strong field spanned between
two color charges moving away from each other.

In this review, we will, however, focus exclusively on the 
concept of universal FFs arising from applying
factorization theorems to hard scattering processes with one
produced hadron and for which 
a proper field-theoretical definition in terms of bi-local 
correlators on the light-cone can be given.
In the case of quarks it reads
\cite{Collins:1981uk,Collins:1981uw,Collins:2011zzd}
\begin{equation} 
\label{eq:operator}
D_q^h(z) = \frac{z}{24\pi} \int dx^- e^{-ip_h^+ x^-/z}
\,\,\text{Tr}\, \Big[ \gamma^+ \langle 0|
\Psi(0) {\cal{P}} |h(p_h^+)X\rangle \langle h(p_h^+)X| 
{\cal{P^{\prime}}} \bar{\Psi}(x)|0\rangle \Big]
\end{equation}
and similarly for antiquarks and gluons.
${\cal{P}}$ and ${\cal{P^{\prime}}}$ denote the necessary 
path-ordered line integrals of the QCD color potential $A$ (gauge-links)
to render the definition gauge-invariant. 
The standard parton model interpretation of $D_i^h(z)$, 
where $z=p_h^+/k_i^+$ is the light-cone momentum fraction taken by
the produced hadron $h(p_h^+)$ from a parton $i(k_i^+)$, is recovered in light-cone $A^+=0$ gauge where 
${\cal{P}}={\cal{P^{\prime}}}=1$.
Because of the observed hadron $h$ in the final state
and the corresponding lack of a complete set of states in Eq.\ (\ref{eq:operator}),
a local operator product expansion (OPE) does not apply for FFs.
For fully inclusive PDFs,
the OPE is the basis for first principle numerical 
computations of some of their integer (Mellin) moments 
within lattice QCD methods \cite{Constantinou:2020hdm}.
Similar calculations cannot be pursued for FFs
but recently one has began to utilize other tools such
large-momentum effective theories to compute light-cone correlators
directly from first principles \cite{Ji:2013dva,Ji:2020ect}.
Perhaps such techniques come to fruition also for FFs in the future.

In the following, we have to limit ourselves to integrated/collinear FFs as
outlined above and defined through Eq.\ (\ref{eq:operator}). 
There are several phenomenologically relevant generalizations of this concept
\cite{Metz:2016swz,Boussarie:2023izj}.
Most importantly, one can restore the transverse momentum dependence
of FFs by undoing the respective integrations in
their definition (\ref{eq:operator}). In addition,
information on the spin of the fragmenting (anti)quark or gluon
can be added.
This leads to a rich suite of novel observables and transverse momentum
dependent (TMD) FFs where the structure of the
gauge links and the underlying TMD factorization theorems become 
much more intricate than in the collinear case \cite{Collins:2011zzd}. 
It is also possible to generalize the definition\ (\ref{eq:operator}) to
include two nearby measured hadrons instead of one, leading to
the concept of di-hadron FFs \cite{Bianconi:1999cd}.

Quite similar to PDFs, performing the integrals in 
Eq.\ (\ref{eq:operator}) leads to ultraviolet divergencies
which have to be removed by some renormalization procedure.
Likewise, various types of singularities appear in computations
of partonic cross sections beyond the leading order (LO) approximation,
for instance, due to soft and/or collinear gluon radiation.
The factorization procedure outlined above introduces arbitrary scales,
here, for simplicity, all denoted as $\mu$,
of the order of the hard scale to separate perturbative from the 
non-perturbative regimes in a theoretical calculation. In this way, the 
FFs $D_i^h(z,\mu)$, like PDFs and the strong coupling $\alpha_s(\mu)$, acquire a scale-dependence which is
absent in the naive parton model where strict scaling of PDFs and FFs is predicted.
When combined with a similar dependence on $\mu$ in the 
hard scattering coefficient functions, physical cross sections are, as they should,
independent of $\mu$, but only up to the perturbative order considered in the calculation. 
By pushing the series expansion in $\alpha_S$ further and
further, one can improve theoretical estimates from LO to 
next-to-LO (NLO) to next-to-NLO order (NNLO) accuracy, 
the present state-of-the-art for most processes involving FFs as we shall discuss below.
In turn, the residual dependence on $\mu$, a major source of theoretical
uncertainties in a cross section estimate, is expected to diminish.

The scale dependence of the FFs $D_i^h(z,\mu)$
is calculable in pQCD through renormalization
group or evolution equations (RGEs) whose kernels, the splitting functions or
anomalous dimensions, are currently known up to NNLO accuracy
\cite{Mitov:2006ic,Moch:2007tx,Almasy:2011eq,Chen:2020uvt}.
The evolution has to stop at some hadronic, bound-state like scale
$\mu_0= \mathcal{O}(1\,\mathrm{GeV})<\mu$
though, where the non-perturbative $D_i^h(z,\mu_0)$ have to be
specified as an initial condition to the RGEs. This is usually achieved by 
global QCD analyses of all or most of the available data 
that can help to constrain the input FFs $D_i^h(z,\mu_0)$. 
In this context, an important but challenging task 
is to propagate the experimental uncertainties from 
different experiments to the extracted set of FFs in a faithful manner.

In the subsequent chapters we will elaborate on all phenomenological
aspects of integrated, collinear FFs in more detail,
comprising their scale evolution, constraints from sum rules,  
and the theory of relevant one-hadron inclusive and related 
processes including
all-order resummations in Chapter \ref{chap2:theory}.
Strategies to extract FFs and their uncertainties 
for pions, kaons, and unidentified charged hadrons
from world data, collected in Chapter \ref{chap3:data},
through global QCD analyses along with the latest, 
state-of-the art results are reviewed in Chapter \ref{chap4:fits}.

\section{Working with fragmentation functions within pQCD}\label{chap2:theory}
%
In this chapter, we review the current status of perturbative QCD calculations
pertinent to the scale evolution of FFs and the processes most crucial
for current determinations of FFs from data, namely SIA, semi-inclusive DIS (SIDIS),
and single-inclusive hadron production in proton-proton collisions.
We also touch upon potentially interesting less
inclusive probes, sum rules obeyed by FFs, and possible improvements 
beyond fixed-order (FO) perturbation theory 
from all-order resummations of potentially large logarithmic 
corrections endangering the stability of FO results.
%
\subsection{Scale dependence of fragmentation functions}
\label{sec:scaledep}
\subsubsection{The time-like evolution equations}
\label{subsec:evol}
%
The dependence of the $D_i^h(z,\mu)$
on the arbitrary factorization scale $\mu$ is governed
by a set of $2N_f+1$ coupled integro-differential RGEs
\cite{Gribov:1972ri,Gribov:1972rt,Lipatov:1974qm,Dokshitzer:1977sg,Altarelli:1977zs}
\begin{equation}\label{eq:evolution}
\frac{\partial}{\partial\ln\mu^2}D^h_i(z,\mu)= \sum_j P^T_{ji}(z,\mu)
\otimes D^h_j(z,\mu),\,\text{where}\;i,j=q,\bar{q},g,
\end{equation}
for the $N_f$ active, i.e., effectively massless quark and antiquark flavors and the gluon.
The symbol $\otimes$ denotes the standard convolution integral
\begin{equation}\label{eq:convolution}
f(z)\otimes g(z)\equiv \int_0^1dx\int_0^1dy \, f(x)\, g(y)\,\delta(z-xy)\, .
\end{equation}
The $j\to i$ time-like splitting functions 
can be calculated perturbatively as a series in $a_s\equiv \alpha_s(\mu)/(4\pi)$,
\begin{equation}\label{eq:splitting}
P^{T}_{ji}(z,\mu) = a_s P_{ji}^{T,(0)}(z) + 
a_s^2 P_{ji}^{T,(1)}(z) +  a_s^3 P_{ji}^{T,(2)}(z) + \ldots \,,
\end{equation}
and are currently known up to NNLO accuracy
\cite{Mitov:2006ic,Moch:2007tx,Almasy:2011eq,Chen:2020uvt}, i.e., including
terms of ${\cal O}(a_s^3)$ in Eq.~(\ref{eq:splitting}), 
as is the case for their space-like counterparts $P^{S}_{ij}(x,\mu)$ 
\cite{Moch:2004pa,Vogt:2004mw}
needed for the analogous scale evolution of PDFs.

It is worthwhile to briefly recall how these calculations of the $P^{T}_{ji}$ played out.
At LO, the time-like and space-like splitting functions are actually equal, 
$P_{ji}^{T,(0)}(\xi)=P_{ji}^{S,(0)}(\xi)$, in their respective physical
domains $\xi<1$ \cite{Gribov:1972ri,Gribov:1972rt,Lipatov:1974qm,Dokshitzer:1977sg,Altarelli:1977zs,
Owens:1978qz,Georgi:1977mg,Uematsu:1978yw}, a manifestation of the 
Gribov-Lipatov reciprocity relation \cite{Gribov:1972ri,Gribov:1972rt}.
Time-like and space-like quantities at LO are also related by analytical
continuation or the Drell-Yan-Levy rule \cite{Drell:1969jm,Drell:1969wb,Drell:1969wd},
which schematically reads
\begin{equation} \label{eq:acrule}
P_{ji}^{T,(0)}(z)= z\, \mathcal{AC}\left[P_{ij}^{S,(0)}(x=\frac{1}{z})\right]  \,. 
\end{equation}
The operation $\mathcal{AC}$ not only analytically continues 
the $P_{ji}^{S,(0)}(x)$ to $x\to\frac{1}{z}>1$ but also appropriately 
adjusts the color factors and an overall sign if needed. The factor $z$ 
in Eq.~(\ref{eq:acrule}) stems from the phase-space of the observed parton in SIA.
Since $\mathcal{AC}$ relies on symmetries of the relevant tree-level diagrams under 
crossing, one does not expect it to hold in general beyond the LO 
due to the need to regulate, renormalize, and factorize various types of divergencies 
in intermediate stages of the calculation.
Here, one usually adopts $n$-dimensional regularization \cite{tHooft:1972tcz} and
the $\overline{\mathrm{MS}}$ scheme \cite{Bardeen:1978yd}.
At NLO accuracy the breaking of Eq.~(\ref{eq:acrule}) can be 
completely understood from kinematical considerations though, and the $\mathcal{AC}$ rule
was used to derive the $P_{ji}^{T,(1)}(z)$ in Eq.~(\ref{eq:splitting})
from their space-like counterparts 
\cite{Curci:1980uw,Furmanski:1980cm,Floratos:1981hs,Munehisa:1981ke,Stratmann:1996hn,Blumlein:2000wh}. 
It should be noted that, apart from $P_{gg}^{T,(1)}(z)$, 
the NLO splitting functions have been also extracted directly 
from jet production in SIA \cite{Kalinowski:1980ju,Kalinowski:1980wea}.
Powerful checks on the obtained results for the 
$P_{ji}^{T,(1)}(z)$, and likewise in case of $P_{ij}^{S,(1)}(x)$,
come from sum rules, see Sec.~\ref{subsec:resum}, 
and a conjectured relation that connects 
all four splitting functions in a remarkably simple way,
$P_{qq}^{T,(1)}+P_{gq}^{T,(1)}-P_{qg}^{T,(1)}-P_{gg}^{T,(1)}=0$,
in the limit of an $\mathcal{N}=1$ supersymmetric Yang-Mills theory
\cite{Dokshitzer:1977sg}, i.e.,
when all QCD color factors are identified $(C_A=C_F=2T_F=N_C)$.
However, for the relation to be satisfied
one needs to perform a factorization scheme transformation \cite{Stratmann:1996hn}
from $\overline{\mathrm{MS}}$ dimensional regularization 
to dimensional reduction \cite{Siegel:1979wq} since only the latter respects supersymmetry.
At NNLO, the application of the $\mathcal{AC}$ rule to the
$P_{ji}^{T,(2)}(z)$ is even more involved and one is left with a small
uncertainty in $P_{qg}^{T,(2)}(z)$ proportional to $C_A-C_F$ that
vanishes in the supersymmetric limit \cite{Almasy:2011eq}.
This ambiguity has been understood and fully resolved only fairly recently
\cite{Chen:2020uvt} in the framework of soft-collinear effective theory (SCET) \cite{Becher:2014oda} 
by analyzing the loss of analyticity for certain contributions.
They also proposed a novel, all-order reciprocity relation for the
Mellin moments of the eigenvalues of the splitting function matrix.
In general, the time-like splitting functions still elude a full direct calculation
both at NLO and NNLO accuracy though.

To better reflect the properties of the (anti)quark to (anti)quark splitting functions,
it is convenient to recast the entire quark sector of Eq.\ (\ref{eq:evolution}) 
into a flavor singlet 
\begin{equation} 
\label{eq:singlet}
D^h_{\Sigma} \equiv \sum_{q}^{N_f}(D^h_{q}+D^h_{\bar q}) \, ,
\end{equation}
which evolves along with the gluon FF $D_g^h$ according to 
\begin{equation}
\label{eq:singletevol}
\frac{d}{d\ln\mu^2}\, \Bigg (\begin{matrix}D^h_{\Sigma} \\[2mm] D^h_g
\end{matrix} \Bigg ) =  \Bigg (\begin{matrix}P^T_{qq} & 2N_f P^T_{gq}\\[2mm] \frac{1}{2N_f}P^T_{qg} & P^T_{gg} \end{matrix} \Bigg ) \otimes \Bigg (\begin{matrix}D^h_{\Sigma} \\[2mm] D^h_g
\end{matrix} \Bigg ) \, ,
\end{equation}
and $2N_f-1$ flavor non-singlet (NS) combinations 
\begin{eqnarray} \label{eq:NSPM}
D^{h,\pm}_{\mathrm{NS},l} \equiv \sum_{i=1}^k (D^h_{q_i}\pm D^h_{\bar q_i}) - k (D^h_{q_k}\pm D^h_{\bar q_k})\,\, \text{where}\;l=k^2-1, k=2,\ldots,N_f\,\, 
\text{and}\,\,
D^h_{\mathrm{NS},v} \equiv \sum_{q}^{N_f}(D^h_{q}-D^h_{\bar q})\,,
\end{eqnarray}
which all evolve independently.
More specifically, upon exploiting the flavor structure
of the splitting functions, i.e., 
$P^{T}_{q_i q_k}=P^{T}_{\bar{q}_i \bar{q}_k}\equiv \delta_{ik} P^{T,v}_{qq}+P^{T,s}_{qq}$
and $P^{T}_{q_i \bar{q}_k}=P^{T}_{\bar{q}_i q_k}\equiv \delta_{ik} P^{T,v}_{q\bar{q}}+P^{T,s}_{q\bar{q}}$,
the two NS flavor asymmetries $D^{h,\pm}_{\mathrm{NS},l}$ and the total valence combination
$D^h_{\mathrm{NS},v}$ in Eq.~(\ref{eq:NSPM}),
evolve with the NS splitting functions 
\begin{equation}
\label{eq:splittingNSPM}
P^{T,\pm}_{\mathrm{NS}}=P^{T,v}_{qq}\pm P^{T,v}_{q\bar q}\,\,\text{and}\,\,
P^{T,v}_{\mathrm{NS}}= P^{T,v}_{qq}-P^{T,v}_{q\bar{q}} + N_f (P^{T,s}_{qq}-P^{T,s}_{q\bar{q}})
\equiv P^{T,-}_{\mathrm{NS}}+P^{T,s}_{\mathrm{NS}} \, ,
\end{equation}
respectively, and the singlet $P_{qq}^T$ in (\ref{eq:singletevol}) obeys
\begin{equation}
\label{eq:splittingS}
P^{T}_{qq} =P^{T,+}_{\mathrm{NS}} +N_f(P^{T,s}_{qq}+P^{T,s}_{q\bar q})
\equiv P^{T,+}_{\mathrm{NS}}+P^{T,ps} \,.
\end{equation}
At LO and NLO accuracy, 
$P^{T,v}_{q\bar q} = P^{T,s}_{\mathrm{NS}} = P^{T,ps} = 0$ holds, whereas at NLO only $P^{T,s}_{\mathrm{NS}}=0$. Consequently, all three independently
evolving NS quark combinations appear for the first time at NNLO accuracy.
A non-vanishing $P^{T,s}_{\mathrm{NS}}$ can lead to a 
perturbatively generated, albeit small strange-quark asymmetry for FFs, i.e.,
$D_s^{h}(z,\mu)-D_{\bar s}^{h}(z,\mu)\neq 0$,
even if $D_s^{h}=D_{\bar{s}}^{h}$ holds at the input scale. 
Since this effect is driven by a non-zero
$D^h_{\mathrm{NS},v}$ in Eq.~(\ref{eq:NSPM}), 
its magnitude can be further influenced by assumptions
based on isospin and charge conjugation symmetries, see Sec.~\ref{subsec:symmetries}.
We note that the four-loop NS space- and time-like splitting functions have been
computed recently \cite{Gehrmann:2026qbl}.

\subsubsection{How to solve the time-like RGEs and available numerical implementations}
\label{subsec:solving-rge}
In general, the structure and solutions of the space-like and time-like evolution equations are
completely analogous apart from replacing PDFs by FFs and the kernels $P^{S}_{ij}$ by $P^{T}_{ji}$. 
Due to the $2\times 2$ matrix-valued nature of Eq.~(\ref{eq:singletevol}), 
no unique closed solution of the integro-differential RGEs exits beyond the LO approximation, 
and one has to resort to some appropriate approximations and/or numerical methods as we shall discuss below. 
Since the three distinct NS combinations in Eq.~(\ref{eq:NSPM}) evolve
independently of each other, their respective scalar RGEs are much more straightforward to solve,
so we do not consider them here in detail.
Once the singlet and NS RGEs are solved, the individual $D_q^h(z,\mu)$ and $D_{\bar{q}}^h(z,\mu)$ can
be easily recovered from the evolved $2N_f$ linear combinations in Eqs.~(\ref{eq:singlet}) and (\ref{eq:NSPM}).

In the direct or momentum space approach one solves Eq.~(\ref{eq:singletevol}) sequentially by
adopting advanced Runge-Kutta algorithms to step through the energy scale from the initial $\mu_0$
to some $\mu>\mu_0$ relevant for computing physical observables.
To this end, the evolution kernels and the FFs (or likewise PDFs) are sampled on a grid of points in $z$, 
guided by a suitable set of polynomial basis functions, that divides the desired range of $z$ 
into sufficiently small sub-intervals. 
In this way, one not only obtains fast and stable numerical results for the FFs
$D^h_i(z,\mu)$, evolved from some, in principle, arbitrary input $z$-shape $D^h_i(z,\mu_0)$
but the same discretization technique in $z$ can be utilized to facilitate
time-consuming convolutions $\otimes$ of FFs (and PDFs) with partonic subprocess 
cross sections to arrive at physical cross sections, see Sec.~\ref{sec:processes} below. 
Publicly available codes comprise
\texttt{QCDNUM} \cite{Botje:2010ay,Botje:2016wbq},
\texttt{Hoppet} \cite{Karlberg:2025hxk}, \texttt{Apfel++} \cite{Bertone:2017gds}, 
and \texttt{APPLGrid} \cite{Carli:2010rw}, where the latter 
tool allows one to pre-compute the necessary grids of partonic short-distance
cross sections. It should be noted that out of the box \texttt{Hoppet} solely
provides the evolution of PDFs and has to be adapted accordingly to evolve also FFs,
see, e.g.~\cite{Gao:2024dbv}.

A mathematically more appealing approach,
that also provides deeper insight into the properties of the RGE (\ref{eq:singletevol}) 
and its solution, is to first turn the integro-differential evolution equations 
into ordinary differential equations before solving them analytically.
This is achieved by means of an integral transform to complex Mellin $N$ moment space.
In general, the Mellin transform $f(N)$ of a function $f(z)$ and its inverse
are defined by
\begin{equation}
\label{eq:mellin}
f(N)=\mathcal{M}[f(z)]=\int_0^1dz\,z^{N-1}f(z) \,\,\text{and}\,\,
f(z) = \frac{1}{2\pi i}\int_{{\cal C}_N} dN\, z^{-N} f(N)\,,
\end{equation}
respectively, where ${\cal C}_N$ is some suitably chosen contour in the complex $N$ plane 
with all singularities of the function $f(N)$ to its left, see, e.g., 
\cite{Vogt:2004ns,Anderle:2015lqa} for further details.

The Mellin-transformed matrix-valued singlet evolution equation~(\ref{eq:singletevol}) 
reads after a convenient change of variables from $\ln \mu^2$ to $a_s(\mu)$ 
\begin{equation}
\label{eq:mtevolution}
\frac{\partial \boldsymbol D^h(N,a_s)}{\partial a_s}=-\frac{1}{a_s}\bigg[\boldsymbol R_0(N)+\sum^\infty_{k=1}a_s^k\boldsymbol R_k(N)\bigg]\boldsymbol D^h(N,a_s)\,\,\text{where}\,\,
\boldsymbol D^h(N,a_s)\equiv
\left(\begin{matrix}
D^h_{\Sigma}(N,a_s) \\D^h_g(N,a_s)
\end{matrix} \right )\,\,\text{and}\,\,
\boldsymbol R_k \equiv \frac{1}{\beta_0}\boldsymbol P^{T,(k)}-\sum^k_{i=1}b_i\,\boldsymbol R_{k-i}\,.
\end{equation}
The non-commutative matrices $\boldsymbol R_k$ at order $k$ are defined recursively in terms of the
perturbative expansion of the time-like splitting functions $\boldsymbol P^{T}$ in $N$ space
and $b_i \equiv \beta_i/\beta_0$. The latter contain the coefficients $\beta_i$ 
of the QCD $\beta$-function, currently known up to five-loop order \cite{Baikov:2016tgj},
that govern the running of the strong coupling $a_s(\mu)$ according to
\begin{equation}
\label{eq:alpha-rge}
\frac{\partial a_s(\mu)}{\partial \ln\mu^2}= \beta(a_s) = -\sum_{k=0}^{\infty} a_s^{k+2} \beta_k\,.
\end{equation}
The RGE (\ref{eq:alpha-rge}) leads to an implicit equation for $a_s(\mu)$ in terms of its initial value
$a_0\equiv a_s(\mu_0)$ and is, hence, best solved numerically. 

Beyond the LO, Eq.~(\ref{eq:mtevolution}) can be only solved as an
expansion around its analytical LO ($k=0$) solution, 
$\boldsymbol D^h(N,a_s)=(a_s/a_0)^{-\boldsymbol R_0(N)} \boldsymbol D^h(N,a_0)$, 
where the non-perturbative input $\boldsymbol D^h(N,a_0)$ is specified from a fit to data.
This expansion can be expressed as
\begin{eqnarray}
\label{eq:mtgeneralsolution}  
\boldsymbol D^h(N,a_s)=\bigg[1+\sum^\infty_{k=1}a_s^k\,\boldsymbol U_k(N)\bigg]
\bigg(\frac{a_s}{a_0}\bigg)^{-\boldsymbol R_0(N)} 
\bigg[1+\sum^\infty_{k=1}a_0^k\,\boldsymbol U_k(N)\bigg]^{-1}\boldsymbol D^h(N,a_0)\,\,\text{where}\,\,
[\boldsymbol U_k,\boldsymbol R_0]=\boldsymbol R_k+\sum_{i=1}^{k-1}\boldsymbol R_{k-i}\boldsymbol U_i+k\boldsymbol U_k
\end{eqnarray}
recursively defines the evolution matrices $\boldsymbol U_k$ through a chain of
commutation relations. A similar but simpler result also 
holds in the NS case, see, for instance,
\cite{Vogt:2004ns} for details.
More specifically, at any given order N$^\text m$LO one can now define several fully 
equivalent solutions based on Eq.~(\ref{eq:mtgeneralsolution}) that differ only in 
sub-leading higher order terms $\mathcal O (a_s^{m+1})$.
After expanding Eq.~(\ref{eq:mtgeneralsolution}) in $a_s$, the frequently used 
\emph{truncated solution} strictly keeps only terms up to $a_s^m$ in both $\boldsymbol R_k$ 
and $\boldsymbol U_k$ and solves Eq.~(\ref{eq:mtevolution}) not literally but only in the
sense of a power series.
Given the iterative nature of the $\boldsymbol R_k$, one may alternatively calculate 
the $\boldsymbol R_k$ and, hence, the $\boldsymbol U_k$, for any $k>m$ from the known 
results for $\boldsymbol P^{T,(k)}$ and $\beta_k$ up to $k=m$.
All higher order contributions with $k>m$ are simply set to zero.
The so defined \emph{iterated solution} closely mimics the results that would be obtained 
by numerically solving Eq.~(\ref{eq:singletevol}) directly in $z$-space as outlined above.
It should be stressed that both choices are equally valid as they only 
differ at $\mathcal O (a_s^{m+1})$.
Any numerical differences in the resulting FFs, albeit small, should be treated 
as part of the theoretical uncertainties in global QCD analyses of FFs (and PDFs).
In ref.~\cite{Anderle:2015lqa} it was estimated that in the $z$-range relevant for the 
extraction of FFs from data, see Chap.~\ref{chap4:fits},
the differences between the truncated and iterated solutions 
amount to less than $1\%$ ($2\%$) at NNLO (NLO) accuracy.

Publicly available codes based on Mellin space comprise
\texttt{Pegasus} \cite{Vogt:2004ns},
\texttt{MELA} \cite{Bertone:2015cwa}, and
\texttt{EKO} \cite{Candido:2022tld}, the latter, being  written in \texttt{python}, 
is the most modern and versatile tool that allows for very flexible non-perturbative
input distributions. It can be easily interfaced with tools like 
\texttt{APPLGrid} or \texttt{PineAPPL} \cite{Jezo:2026adf} to compute physical cross sections.
Instead of performing these convolutions in $z$-space, one can also pre-compute
the Mellin moments of the relevant partonic cross sections 
on a suitable grid of complex $N$ values  \cite{Graudenz:1995sk,Kosower:1997vj,Stratmann:2001pb} and
multiply them with the corresponding moments of the evolved FFs. The physical cross section
is then obtained by a fast numerical inverse Mellin transform based on a Gaussian quadrature 
on the predefined support points in $N$.
All Mellin codes are numerically very efficient and deliver precise results.
Like for \texttt{Hoppet}, \texttt{Pegasus} first has to be extended from PDF to FF evolution
\cite{Anderle:2015lqa}.

An important comment is in order at this point. For simplicity we have assumed 
so far that there is only one common scale $\mu$. 
In fact, since these solely theoretical scales can be chosen arbitrarily as measurable
quantities should not depend on them, different choices can be made in the factorization of FFs 
(and PDFs for that matter) and the renormalization of the strong coupling $a_s$, 
usually labeled as $\mu_f$ and $\mu_r$, respectively.
Expressions for the general case $\mu_f\neq\mu_r$, usually implemented
in the evolution codes, can be derived by first replacing all couplings with
\begin{equation}
a_s(\mu_f)=a_s(\mu_r)\left[1+a_s(\mu_r)\,\beta_0\ln{\frac{\mu_r^2}{\mu_f^2}}+\mathcal{O}(a_s^2)\right]
\end{equation}
up to sufficiently high order, followed by re-expanding the splitting functions 
and partonic cross sections in terms of $a_s(\mu_r)$. This procedure 
yields a plethora of additional logarithms of the type $\log(\mu_r^2/\mu_f^2)$ 
which motivates the common, simplifying choice $\mu_f=\mu_r=\mu$ in 
most phenomenological analyses. 
Nevertheless, varying $\mu_f$ and $\mu_r$ independently
within a reasonable range around the hard scale
characterizing the physical process, 
typically by a factor of two, likely gives a more
faithful estimate of the theoretical uncertainties
due to the truncation of the perturbative series.
It should be noted that the switch of variables in
(\ref{eq:mtevolution}) is only permitted as long as 
$\mu_r$ is a multiple of $\mu_f=\mu$.
%
\subsubsection{Treating heavy quark flavors in the evolution}
\label{subsec:heavyquarks}
The RGEs in Secs.~\ref{subsec:evol} and \ref{subsec:solving-rge}
that govern the scale dependence of FFs and $a_s$ are set up
for massless partons. Thus, so far we can only deal with the
case of a fixed number $N_f$ of active (massless) quark flavors and the gluon 
in the evolution, the so-called fixed flavor number scheme (FFNS). 
For this purpose, a quark is defined as light (i.e., massless) or heavy according
to whether its mass is much less or greater than the 
relevant scale $\mu$ \cite{Collins:1978wz}. 
Typically, the boundary conditions for the RGEs
(\ref{eq:mtevolution}) and (\ref{eq:alpha-rge})
are provided at scales $\mu_0\simeq \mathcal{O}(1\, \mathrm{GeV})$,
such that only the $u$, $d$ and $s$ (anti)quarks are 
effectively massless, i.e.~$N_f=3$.

However, FFs have applications in processes where the 
relevant scales, e.g., $\sqrt{S}$ or the transverse momentum
$p_T$ of the observed hadron, range from a few $\mathrm{GeV}$ up to 
100 or more $\mathrm{GeV}$, see Sec.~\ref{sec:processes} and Chap.~\ref{chap3:data}.
Therefore, one needs a prescription of how to cross the threshold from $N_f$
to $N_f+1$ flavors, i.e., a variable flavor number scheme (VFNS) 
that one-by-one incorporates the heavy (anti)quarks (HQs), 
$Q=c,b$ and $\bar{Q}=\bar{c},\bar{b}$, of mass $m_Q$ at 
large enough scales $\mu\gtrsim m_Q$, where they can be approximately
treated as massless partons. Below this matching scale, the HQ is
decoupled from the theory.

To this end, matching conditions for the HQ-to-hadron FFs 
have been computed perturbatively up to NLO accuracy 
in the $\overline{\text{MS}}$ scheme by
exploiting the large scale $m_Q$ and read \cite{Cacciari:2005ry}
\begin{equation}
D_Q^h(z,\mu) = D_{\bar{Q}}^h(z,\mu) = \frac{\alpha_s(\mu)}{2\pi}
\int_z^1 \frac{dy}{y} D_g^h\left(\frac{z}{y},\mu\right)
C_F \frac{1+(1-y)^2}{y} \left[\ln \frac{\mu^2}{m_Q^2} -1-2\ln y  \right]\,,
\end{equation}
which is readily transformed into Mellin space as well. For the standard choice
$\mu=m_Q$ for the $N_f\to N_f+1$ matching scale, 
the light quarks and gluon FFs are continuous at NLO.
Corresponding expressions at NNLO accuracy are not known yet.
These matching conditions should be used in fits that generate 
FFs for charm and bottom quarks purely dynamically, i.e., $D_Q^h$ is
fully determined by the FFs of the gluon and the light quarks in the 
$N_f$-scheme.
Of course, upon crossing a HQ threshold, the running of the strong coupling needs
to be adjusted as well as it also depends on the number of active 
massless flavors $N_f$ through $\beta(a_s)$ in (\ref{eq:alpha-rge}).
Here, the needed decoupling relations are known up to $\mathcal{O}(\alpha_s^4)$ 
\cite{Schroder:2005hy,Chetyrkin:2005ia}.

In practice, however, most analyses of FFs performed so far 
have adopted a different approach to implement HQ FFs through a VFNS.
Once the scale $\mu$ in the evolution crosses $m_{Q}$, a new non-perturbative boundary condition 
$D^h_Q(z,\mu) = D^h_{\bar{Q}}(z,\mu)$ is introduced discontinuously and $N_f\to N_f+1$.
The new free parameters are in turn determined 
by a fit to suitably flavor-tagged data taken at scales $\mu \gg m_Q$, 
see Chap.~\ref{chap3:data} and \ref{chap4:fits}.
The evolution codes listed in Sec.~\ref{subsec:solving-rge} are usually 
capable of implementing different schemes for HQ FFs.

While such a zero-mass VFNS (ZM-VFNS) appears to be sound if applied at scales $\mu \gg m_Q$
where HQs are effectively massless, potential shortcomings may arise if HQ FFs are
used in calculations when $m_Q$ is not necessarily
negligible with respect to the relevant large momentum scale characterizing the data. 
Close to the HQ threshold $\mu\simeq m_Q$ it is certainly much more suitable 
to fully retain the HQ masses in calculations of hard scattering cross sections,
but then potentially large logarithms $\ln (\mu/m_Q)$ may spoil the
convergence of the perturbative series once $\mu \gg m_Q$.
Resumming these logarithms to all orders leads one back to the ZM-VFNS.
Therefore, in PDF fits, various general mass VFNS (GM-VFNS) have been introduced to
smoothly interpolate the fully massive results to those obtained in a ZM-VFNS at some
intermediate scales $\mu$. The implementation of such a GM-VFNS is, however, not unique 
and introduces assumptions that have to be accounted for as part of the
theoretical uncertainties, see, e.g.~\cite{Accardi:2016ndt}.
Apart from exploratory studies \cite{Epele:2016gup,Epele:2018ewr}, no GM-VFNS has been fully 
developed or used yet in the context of global QCD analyses of FFs for light hadrons.

This situation is different though if one considers the fragmentation of HQs into
heavy-quark-flavored hadrons such as $D$- and $B$-mesons, whose rich phenomenology is beyond the scope of our review focusing exclusively on FFs for light hadrons.
The interested reader may want to consult \cite{Bonino:2023icn,Fickinger:2016rfd,Cacciari:2012ny,Kniehl:2011bk}
and references therein as a starting point for this particular aspect of hadronization.

\subsubsection{Sum rules, multiplicities, all-order resummations, and all that}
\label{subsec:resum}
%
In this section, we briefly discuss some further properties of FFs and possible
theoretical refinements for calculating their scale dependence by including all-order
resummations of potentially large logarithms or information on the polarization of
the fragmenting partons and the observed hadron.

We start with sum rules, which encode conservation laws in QCD,
and recall that in case of PDFs the momentum and flavor/quantum number sum rules,
$\sum_i \int_0^1 x  f_i^h(x,\mu) \, dx = 1$ and 
$\int_0^1 [f_u^p(x,\mu)-f_{\bar{u}}^p(x,\mu)]\,  dx = 2$,
$\int_0^1 [f_d^p(x,\mu)-f_{\bar{d}}^p(x,\mu)]\,  dx = 1$,
respectively, yield valuable constraints on the parameter space describing PDFs.
The validity of the sum rules for all scales $\mu$ is embedded into the
properties of the space-like splitting functions $P_{ij}^{S}(x)$
in their perturbative expansion, e.g., by enforcing that
$\int_0^1 x\,[P_{qq}^{S,(k)}(x)+P_{gq}^{S,(k)}(x)] dx=0$ to all orders $k$ .

Analogous sum rules hold in the time-like case for FFs \cite{Collins:1981uw}. 
The fundamental momentum and quantum number/charge sum rules 
now read for each parton flavor $i$ and for arbitrary scales $\mu$
\begin{equation}
\label{eq:sumrule-tl}
\sum_h \int_0^1 z D_i^h(z,\mu) \, dz = 1 \,\,\text{and}\,\, 
\sum_h \mathcal{Q}_h \int_0^1 D_i^h(z,\mu) \, dz = \mathcal{Q}_i\,,
\end{equation}
respectively, for a general conserved quantity $\mathcal{Q}$ 
such as the electric charge, isospin, strangeness and so on.
Again, in a conventional factorization scheme like $\overline{\text{MS}}$ 
the sum rules (\ref{eq:sumrule-tl}) are preserved to all orders in QCD thanks to the 
properties of the time-like splitting functions $P_{ji}^{T}(z)$.
A recent theoretical study \cite{Collins:2023cuo} 
has cast some doubt about the strict validity
of the sum rules (\ref{eq:sumrule-tl}) in QCD due
to its confinement property, which leads to subtle issues with
the original proof \cite{Collins:1981uw} when performing the sum over all
hadrons $h$ in Eq.~(\ref{eq:sumrule-tl}). 
However, since one never really has access to the hadronization into \emph{all}
possibly produced hadrons $h$ in phenomenological analyses 
of FFs, none of the constraints in Eq.~(\ref{eq:sumrule-tl}) can
ever be utilized in practice. Nonetheless, in particular the momentum sum
rule still can serve as an important consistency check or upper bound in fits
by ensuring that after adding up all the extracted FFs for a given parton $i$ 
into, say, pions, kaons, and (anti)protons, the sum rule is not already over-saturated.
A numerical study based on recent fits of FFs can be found in \cite{Gao:2024dbv},
see also \cite{Kotlorz:2025xso} for charge sum rules.

We note that a generic problem with analyzing or utilizing moments like 
(\ref{eq:sumrule-tl}) arises from the fact that data always constrain FFs
(or PDFs, for that matter) only in a more or less limited range of $z$ 
such that extrapolations and assumptions are needed to evaluate the integrals
in the entire range of $z$.
In addition, the convergence of the perturbative series in the limits $z\to 0$
and $z\to 1$ might come under attack from an ever increasing set of
potentially large logarithms both in the scale evolution of FFs
and in the hard scattering processes. As a remedy, more and more 
sophisticated theoretical methods for all-order resummations of these contributions 
have been devised based on existing FO results. 
Here we can only touch upon the gist of the matter and have to refer the reader to the literature.

The need for resummations in the context of FFs dates back to the
phenomenological interest in the $z$-integrated single-inclusive 
hadron spectrum in $e^+e^-$ annihilation, which can be related to the $N=1$ Mellin 
moment of FFs, i.e., $\int_0^1 D_i^h(z,\mu) dz$, and yields the energy dependence of 
the multiplicity of a hadron $h$ inside a quark or gluon jet \cite{Dokshitzer:1991wu}.
Data for the underlying energy spectrum are usually plotted in terms of $\xi\equiv -\ln z$
to better emphasize the region of hadrons with low momentum fractions $z$ 
that are expected to dominate the multiplicity.
The shape in $\xi$ exhibits a characteristic and famous
hump-backed plateau (HBP) \cite{Azimov:1985by} that reflects
color coherence and interference of soft gluon radiation in QCD \cite{Dokshitzer:1987nm}.

In FO QCD, the energy dependence of hadron multiplicities and the important
region of hadronization at small $z$ are both ill defined since the 
time-like splitting functions at order $k$ behave like
\begin{equation}
\label{eq:logP}
P_{gi}^{T,(k)} \propto \frac{1}{z} \ln^{2k-a} z\,,\,\text{where } i=q,g,
\end{equation}
in the limit $z\to 0$. Hence, they develop two additional, singular powers of 
$\log^{2k-a}z$ in each order $k$ of perturbation theory,
where $a=0,1,$ and $2$ correspond to the leading logarithmic (LL),
next-to-LL (NLL), and next-to-NLL (NNLL) contribution, respectively, 
all of which can spoil the convergence of the expansion in $a_s$ 
even when the strong coupling is numerically very small. 
The framework to resum these logarithms to all orders in $a_s$ to
mitigate their singular small-$z$ behavior has been known for some time
\cite{Mueller:1981ex,Mueller:1982cq,Mueller:1983js,Mueller:1983cq}.
Most conveniently, these calculations are carried out in Mellin $N$ space, where
the small-$z$ logarithms in Eqs.~(\ref{eq:logP}) transform to
\begin{equation}
\mathcal{M}\left[\frac{\ln^{2k-a}z}{z} \right] = (-1)^a \frac{(2k-a)!}{(N-1)^{2k-a+1}}\,,
\end{equation}
making the singularity in $N=1$ explicit.
An overview of the systematics of $1/(N-1)$ divergences turning up in the time-like scale
evolution of FFs up to NNLL is given in Table~\ref{tab:nnll}. 
\begin{table}[ht!]
\centering
\TBL{\caption{LL, NLL, and NNLL contributions to the time-like splitting functions 
\mbox{$P_{ji}^{T} = \sum_k a_s^{k+1} P_{ji}^{T,(k)}$} at order $k$.  
As indicated in the bottom row, the entries start from $k=0$ (LO)
or $k=2$ (NNLO).}\label{tab:nnll}}
{
\begin{tabular}{lcccc}
\toprule
 & $P_{gg}^{T,(k)}$ & $P_{gq}^{T,(k)}$ & $P_{qq}^{T,(k)}$  &  $P_{qg}^{T,(k)}$ \\[1mm]\hline
LL \rule[0mm]{0mm}{4mm}  & $(N-1)^{-1-2k}$ & $(N-1)^{-1-2k}$ & -- & -- \\[1mm]
NLL & $(N-1)^{-2k}$ & $(N-1)^{-2k}$ & $(N-1)^{-2k}$ & $(N-1)^{-2k}$ \\[1mm]
NNLL & $(N-1)^{1-2k}$ & $(N-1)^{1-2k}$ & $(N-1)^{1-2k}$ & $(N-1)^{1-2k}$ \\[1mm] \hline
\rule[0mm]{0mm}{4mm} & $k\ge0$ & $k\ge0$ &$k\ge2$& $k\ge2$\\
\botrule
\end{tabular}}{}
\end{table}
Note that no LL contributions appear in the quark sector and that
the sub-leading small-$z$ contributions to $P^T_{qq}$, and $P^T_{qg}$ 
at LO and NLO accuracy $(k=0,1)$ are not contained in the table. 
In general, the knowledge of a FO result up to $\mathrm{N}^m\mathrm{LO}$ accuracy fully determines
the first $m+1$ towers of logarithms to all orders. Hence, the scale evolution of FFs can currently
be carried out up to NNLO+NNLL accuracy. Explicit expressions
based on general considerations on the structure of all-order mass factorization can be found in
\cite{Vogt:2011jv, Kom:2012hd}. 
For practical applications it is important to properly match the resummed towers of
logarithms to the full $z$-dependence of the splitting functions to avoid double counting 
and to ensure the validity of the momentum sum rule. 
Details on the implementation of the singlet evolution (\ref{eq:mtevolution}) at NNLO+NNLL accuracy 
in Mellin space, that includes matching and also restores the full dependence on the scales $\mu_f$ and $\mu_r$, 
respectively, can be found in ref.~\cite{Anderle:2016czy}.
Given that the logarithmic contributions to the NS splitting function are sub-leading up to 
NNLL accuracy \cite{Kom:2012hd}, no small-$z$ effects have to be considered in this case and
the usual FO evolution equations apply.

In the context of hadron multiplicities, the modified LL approximation (MLLA) has historically played an important
role \cite{Dokshitzer:1991wu}. 
It resums only the LL plus a few selected sub-leading contributions but gives a reasonable description
of the energy dependence of the hadron spectrum in terms of $\xi$ and the HBP.
To this end one often adopts a distorted Gaussian distribution for the FFs 
\cite{Fong:1990nt} and computes the peak position of the HBP, its width, skewness, and kurtosis
with essentially only the QCD scale $\Lambda_{\text{QCD}}$ as a free parameter. Fits to
the spectra then even allow for competitive determinations of the strong coupling $\alpha_s$.
Some recent work and further refinements related to the MLLA and hadron multiplicities can be found in
\cite{Bolzoni:2013rsa,Perez-Ramos:2013eba}.

Potentially large logarithmic contributions also appear in the scale evolution of FFs 
in the limit $z\to 1$, which corresponds to $N\to \infty$ in Mellin space. The diagonal
splitting functions $P^T_{qq}$ and $P^T_{gg}$ exhibit a stable form with the single logarithm
$\ln N$ being the most singular term \cite{Korchemsky:1988si,Dokshitzer:2005bf}. 
The off-diagonal entries are suppressed by $1/N$ but develop
a tower structure with a double-logarithmic enhancement of the form $\log^{2k-a}(N)$
analogous to the case $z\to 0$.
Results for $a\le 2$ (NNLL) can be found in \cite{Almasy:2015dyv}. The numerical impact
on the scale evolution of FFs is rather limited though.
The consistency of the resummation procedure at a certain logarithmic accuracy
with the factorization theorem demands, however, to carry out similar resummations also for 
the process-dependent partonic cross sections. Hence, we shall get back to the 
potential impact of all-order resummations also in Sec.~\ref{sec:processes}.

Finally, we note that the scale evolution of FFs outlined so far can be extended further
in several directions. First of all, one can include information on the polarization of the
produced hadron $h$ albeit hard to obtain experimentally. 
This is easily achieved by replacing the unpolarized time-like
splitting functions by their corresponding longitudinal or transverse spin-dependent 
counterparts $\Delta P^T_{ji}(z,\mu)$ \cite{Stratmann:1996hn} 
and $\delta P^T_{ji}(z,\mu)$ \cite{Stratmann:2001pt}, respectively, which are
known up to NLO accuracy.
Phenomenological applications are limited so far and have focused on FFs
for $\Lambda$ baryons due to its self-analyzing dominant weak decay $\Lambda\to p \pi^-$
\cite{Perkins:1982xb} that offers the rather unique possibility to reconstruct its spin 
experimentally; see, e.g., \cite{deFlorian:1997zj} for a study.
Secondly, the evolution equations can be generalized to account also for
parton-to-photon FFs $D_i^{\gamma}(z,\mu)$. 
They are of relevance in calculations of prompt photon production cross sections, see, e.g.
\cite{Chen:2022gpk}.
Since the photon can also act as a parton, an additional inhomogeneous piece arises
which makes the scale evolution of $D_i^{\gamma}(z,\mu)$ more elaborate than for
ordinary hadrons. The necessary framework at NLO accuracy can be found in \cite{Gluck:1992zx}.
Alternative sets of photon FFs can be found in \cite{Bourhis:1997yu}. 

\subsection{One-hadron inclusive processes and other less inclusive probes of FFs}
\label{sec:processes}
%
In this section we highlight the main features and the status of perturbative calculations
of the one-hadron inclusive processes most relevant for current determinations of FFs from 
global QCD analyses discussed in Chap.~\ref{chap4:fits} below.
By applying factorization, we show how each process can probe different aspects of FFs (and PDFs).
To this end, it is crucial to perform all necessary subtractions and renormalizations consistently 
in the same scheme, for which the $\overline{\mathrm{MS}}$ scheme \cite{Bardeen:1978yd} is adopted.
Likewise, the knowledge and all-order resummations of large logarithms in the
scale evolution of FFs, outlined in Sec.~\ref{subsec:resum}, alone is not
sufficient to evaluate the hadronic cross sections in the small and large-z region
as large logarithms also appear in the partonic cross sections.
The consistency of the resummation procedure with the factorization theorem demands
the same logarithmic accuracy also in the latter case. Hence, we briefly review the status
of such calculations for one-hadron inclusive processes.
We also comment on power/higher-twist corrections to the perturbative expansion in $a_s$.
One source of purely kinematical origin is the finite mass $m_h$ of the produced hadron $h$
that is neglected in the framework based on factorization. 
In subsection \ref{sec:injet}, we touch upon less inclusive probes of FFs that have not yet  
been routinely used in fits, including hadrons in a jet, hadron pairs, and the 
associated production of a hadron and a vector meson.

\subsubsection{Single-inclusive electron-positron annihilation}
\label{sec:sia-theory}
\begin{figure}[ht!]
\centering
\includegraphics[width=.3\textwidth]{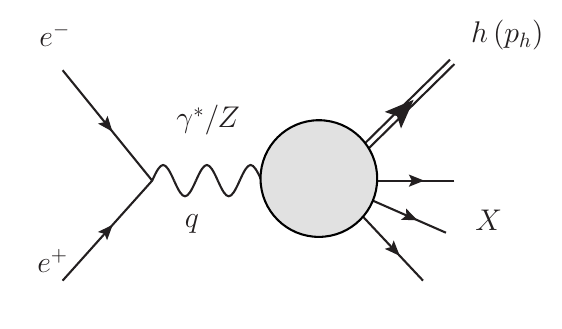}
\caption{Schematic representation of SIA, $e^+e^-\rightarrow \gamma^*/Z\rightarrow h\,X$,
where the grey blob represents all Feynman diagrams contributing to the production of 
an identified hadron $h$ at a given order in pQCD. 
$X$ denotes any experimentally not observed hadronic remnant.}
\label{fig:SIA}
\end{figure}

%
SIA, depicted in Fig.~\ref{fig:SIA}, is the simplest yet the most important process 
pertinent to the notion of FFs \cite{Collins:2011zzd}
and plays a similar definitive role for FFs as DIS does for PDFs.
It proceeds through the exchange of a virtual photon $\gamma^*$ or a $Z$ boson with 
four-momentum $q$. 
Their virtuality squared $Q^2=q\cdot q=S$ is determined by the 
given c.m.s.\ energy $\sqrt{S}$ of the $e^+e^-$ collision.
%
%

The cross section for SIA can depend only 
on the energy $E_h$, conveniently expressed through the energy fraction 
$z=x_E=2E_h/Q=2 p_h\cdot q/Q^2\in[0,1]$, and the production angle $\theta_h$ of the produced hadron 
with four-momentum $p_h$. 
A general tensor decomposition reveals three contributions to $d\sigma^h/dz d\cos\theta_h$
\cite{Mele:1990cw,Nason:1993xx},
two associated with the transverse (T) or longitudinal (L) polarization states of the 
exchanged boson and an asymmetric (A) piece due to the parity-violating nature
of the $Z$ boson that vanishes upon integration over $\cos\theta_h$ which we will 
assume to be performed from now on. Then we can write
\begin{equation}
\label{eq:ee-xsec}
\frac{1}{\sigma_{\mathrm{tot}}}
\frac{d\sigma^h}{dz} = \frac{1}{\sigma_{\mathrm{tot}}} \left[ 
\frac{d\sigma^h_T}{dz}+\frac{d\sigma^h_L}{dz} \right]
\equiv \frac{1} {\sum_q \hat{e}_q^2}
\left[F_T^{h}(z,Q^2) + F_L^{h}(z,Q^2) \right]\,\,
\text{with}\,\,
F_{k}^h \equiv \frac{1}{3\sigma_0} \frac{d\sigma^h_{k}}{dz}
\,\,
\text{and}\,\,k=T,L\,,
\end{equation}
and where $\sigma_0= 4\pi\alpha^2_{\text{em}}/(3 S)$ 
is the point-like QED cross section for $e^+e^-\to \mu^+ \mu^-$.
$\alpha_{\text{em}}$ denotes the electromagnetic fine structure constant 
whose dependence on $\sqrt{S}$ is reviewed in \cite{ParticleDataGroup:2024cfk}.
Explicit formulas for the electroweak quark charges $\hat{e}_q$ can be found, 
e.g., in \cite{Nason:1993xx,Rijken:1996ns}.
The sum in Eq.~(\ref{eq:ee-xsec}) runs over all $N_f$ active massless quark flavors.
As is customary, structure functions $F_{T,L}$ have been introduced in (\ref{eq:ee-xsec})
to draw on the close resemblance to DIS and PDFs \cite{Altarelli:1979kv}.

The SIA cross section (\ref{eq:ee-xsec}) is normalized to the total hadronic
cross section $\sigma_{\mathrm{tot}}$, which up to NNLO accuracy is given by
\cite{Chetyrkin:1979bj,Dine:1979qh,Celmaster:1979xr,Rijken:1996vr}
\begin{eqnarray}\label{eq:nnlotot}
\sigma_{\mathrm{tot}} = 3 \sigma_0 \sum_q \hat{e}_q^2 
\left[ 1+ 3 C_F\,  a_s(\mu_r)  
+ a_s^2(\mu_r) \left(-\frac{3}{2} C_F^2  
+\, C_A C_F \left(-11\ln\frac{Q^2}{\mu_r^2}
- 44 \, \zeta(3) +\frac{123}{2} \right)  
+\, N_f C_F T_f\left[ 4 \ln\frac{Q^2}{\mu_r^2}
+16\,\zeta(3)-22 \right] \right) \right]  \,, 
\end{eqnarray}
with the usual QCD color factors $C_A=3$, $C_F=4/3$, and $T_f=1/2$. 
In this way, integrating Eq.~(\ref{eq:ee-xsec}) over $z$ yields the average
total number or multiplicity of a hadron $h$ in a single SIA event which
is of particular interest phenomenologically, see Sec.~\ref{subsec:resum}.

Applying factorization to SIA, the transverse and longitudinal cross sections
$d\sigma^h_k/dz$ and structure functions $F_k^{h}$ in Eq.~(\ref{eq:ee-xsec}) 
can be expressed as a convolution of appropriate combinations of quark and gluon FFs 
and calculable coefficient functions $\mathbb{C}_{k,l}^{\mathrm{S,NS}}$
\cite{Rijken:1996vr,Rijken:1996ns},
\begin{eqnarray}\label{eq:FTFL}
F_k^h(z,Q^2) =  \left(\sum_q \hat{e}_q^2 \right)
\left[D_\mathrm{S}^h(z,\mu)\otimes 
                              \mathbb{C}_{k,q}^\mathrm{S}\left(z,\frac{Q}{\mu}\right)  
 +   D_g^h\left(z,\mu\right)\otimes 
                            \mathbb{C}_{k,g}^\mathrm{S}\left(z,\frac{Q}{\mu}\right) \right] 
 + \,\sum_q \hat{e}_q^2\,D_{\mathrm{NS},q}^h(z,\mu) \otimes  
                            \mathbb{C}_{k,q}^\mathrm{NS}\left(z,\frac{Q}{\mu}\right) 
\equiv \sum_{l=q,\bar{q},g} D_l^h(z,\mu)\otimes 
                              \mathbb{C}_{k,l}\left(z,\frac{Q}{\mu}\right)\,,
\end{eqnarray}
where $D_\mathrm{S}^h=D_\Sigma^h/N_f$ and $D_\mathrm{NS,q}^h=D_q^h+D_{\bar{q}}^h-D_\mathrm{S}^h$.
For simplicity, we have chosen all scales to be equal, i.e., $\mu=\mu_f=\mu_r$, and $\mu$
should be of the order of the only hard scale $Q$ available in SIA. The general
case $\mu_f\neq\mu_r$ can be recovered as outlined at the end of Sec.~\ref{subsec:solving-rge},
see also \cite{Anderle:2016czy}.
The $\mathbb{C}_{k,l}^{\mathrm{S,NS}}$ can be expanded as series in $a_s$ and are
known up to $\mathcal{O}(a_s^2)$ \cite{Rijken:1996vr,Rijken:1996ns,Mitov:2006wy,Mitov:2006ic},
which is NNLO accuracy for the transverse coefficient functions but formally only NLO for the
$\mathbb{C}_{L,l}^{\mathrm{S,NS}}$ as the latter start to be nonzero only at $\mathcal{O}(a_s)$.
Disregarding any QCD corrections, SIA mainly probes the sum of quark and antiquark FFs
weighted by their appropriate electroweak charges squared. The gluon FF 
appears only as a higher order correction.
We note in passing that the $\mathcal{O}(a_s^2)$ QCD corrections to the parity-violating 
contribution to SIA and for the case where the colliding leptons and the observed hadron
are longitudinally polarized have been computed in \cite{Rijken:1996npa} and 
\cite{Rijken:1997rg,Ravindran:2000rz}, respectively.

Once again, the convolutions in Eq.~(\ref{eq:FTFL}) turn into ordinary products upon
transformation into Mellin $N$ space. Analytic expressions for all known
$\mathbb{C}_{k,l}^{\mathrm{S,NS}}(N,Q/\mu)$ can be found in \cite{Mitov:2006wy,Blumlein:2006rr}
and are readily utilized in conjunction with the N space evolution codes mentioned 
in Sec.~\ref{subsec:solving-rge}.
The singlet coefficient functions 
$\mathbb{C}_{k,l}^{S} = \sum_{n=0} a_s^n \mathbb{C}_{k,l}^{S,(n)}$ exhibit a similar
systematics of double logarithmic enhancements for $z\to 0$ at any given FO $n$
as the splitting functions, see Sec.~\ref{subsec:resum} and Table~\ref{tab:nnll}.
For instance, in the most singular gluon sector one finds that
$\mathbb{C}_{T,g}^{S,(n)} \propto \frac{1}{z} \ln^{2n-1-a}(z)$, where $a=0,1,2$
defines again the  LL, NLL, and NNLL approximation, respectively.
Again, all-order resummations are most conveniently carried out in Mellin space and
the relevant expressions up to NNLL accuracy can be found in \cite{Vogt:2011jv,Kom:2012hd}.
A consistent implementation for SIA, including NNLO+NNLL corrections, properly matched to
the NNLO FO results, for both the scale evolution of the FFs and the SIA coefficient functions
was performed in \cite{Anderle:2016czy}.
While little improvement in the quality of fits was found compared to NNLO results,
the residual scale dependence was significantly reduced in the presence of resummations.

Likewise, large logarithms also appear in the limit $z\to 1$, i.e., $N\to \infty$
in the coefficient functions and have been resummed to at least NNLL accuracy as well 
\cite{Blumlein:2006pj,Moch:2009my,Vogt:2010cv,Almasy:2015dyv}.
Unlike for the splitting functions, see  Sec.~\ref{subsec:resum}, the 
resulting numerical corrections are sizable for the coefficient functions in SIA.
However, terms beyond the NNLL approximation are needed in the threshold expansion
to arrive at reliable and stable resummed results \cite{Almasy:2015dyv}.
Phenomenological studies \cite{Albino:2008fy,Anderle:2012rq}
at NLO+NLL accuracy \cite{Cacciari:2001cw} have found noticeable improvements
in the description of SIA data at the lowest c.m.s.~energies though.

Besides the logarithmic perturbative corrections also non-perturbative
power corrections will ultimately become relevant and need to be addressed.
For one thing, a factorized structure such as in Eq.~(\ref{eq:FTFL}) can only
hold in the presence of a hard scale, say $Q\gg\Lambda_{\mathrm{QCD}}$,
and, hence, up to corrections that are suppressed by inverse powers of $Q$.
More importantly, any perturbative series expansion in QCD is at best
asymptotic for any value of $a_s$ due to divergencies introduced 
by so-called infrared renormalons \cite{Beneke:1998ui,Beneke:2000kc}.
In this way, pQCD is inseparably tied to non-perturbative power corrections 
such that their scaling behavior in $Q$ can be inferred from the renormalon divergence.
Diagrammatically, power or higher-twist corrections arise from multi-parton
correlations that are not included in (\ref{eq:FTFL}).
In case of SIA, it was found that $1/Q$ corrections are present in the longitudinal and 
transverse cross sections, which cancel, however, in the total cross section
\cite{Dasgupta:1996ki,Beneke:1997sr}. This is more severe than the $1/Q^2$ corrections
related to DIS. Power corrections are usually not considered in current global fits
of FFs except for providing some guidance on necessary cuts on the available data
to safely stay in the perturbative regime.

One potentially important kinematical source of power corrections
arises from the finite mass $m_h$ of the produced hadron $h$ that is 
neglected in the factorized framework.
Some experiments quote their data in terms of the scaled three-momentum
$x_p=2|\vec{p}_h|/Q$ of the hadron rather than the energy scaling variable 
$x_E$ in the c.m.s. of the collision. For a finite $m_h$ this leads to the
relation \cite{Nason:1993xx}
\begin{equation}
\label{eq:xp}
x_p=x_E-\frac{2m_h^2}{x_E Q^2} + {\cal{O}}\left(\frac{1}{Q^4}\right)\,,
\end{equation}
which is formally a $1/Q^2$ power correction but the differences between
$x_p$ and $x_E$ are strongly enhanced at small $x_E$ and $\sqrt{S}$, in particular
for heavier hadrons.
Fits of FFs for various hadron species including mass effects and threshold resummations
were performed in \cite{Albino:2008fy}, see also
\cite{Accardi:2014qda} for a similar but more recent analysis.

\subsubsection{Semi-inclusive deep inelastic scattering}
\label{sec:sidis-theory}
\begin{figure}[ht!]
\centering
\includegraphics[width=.3\textwidth]{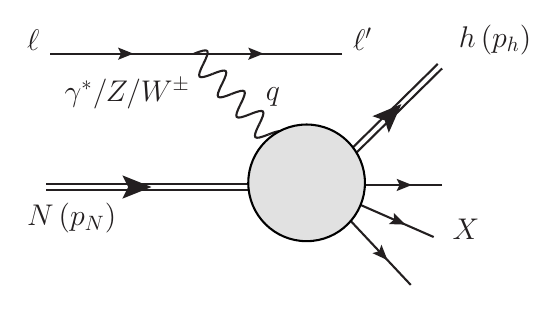}
\caption{Schematic representation of SIDIS, $\ell N \rightarrow \gamma^*/Z/W^{\pm}\rightarrow \ell'\,h\,X$,
where the grey blob represents all Feynman diagrams contributing to the production of 
an identified hadron $h$ at a given order in pQCD.}
\label{fig:SIDIS}
\end{figure}

%
The production of an identified hadron in DIS, or semi-inclusive DIS, represented in 
Fig.~\ref{fig:SIDIS}, complements the constraints on the FFs coming from SIA. 
In a similar fashion to SIA, the cross section for SIDIS, 
$\ell N \rightarrow \gamma^*/Z/W^{\pm}\rightarrow \ell'\,h\,X$, 
can be written in terms of structure functions, 
which in this case involve combinations of both the PDFs of the incoming nucleon $N$ 
and the FFs producing the identified hadron $h$. 
Depending on whether the initial- and final-state leptons $\ell$ and $\ell'$
are the same, the process can be 
mediated by neutral-current (NC) or charged-current (CC) interactions
through the exchange of an appropriate electroweak gauge boson 
with virtuality $Q^2=-q^2$ where $q=p_{\ell}-p_{\ell'}$.
%
%

While the description of the SIDIS process is more convoluted due to the presence of an
initial-state hadron and its associated PDFs, this can be turned into an advantage as it 
allows one to probe and constrain different combinations of FFs than those already present in SIA, 
thus greatly improving the flavor and quark/antiquark separation of FFs in global analyses. 
Hadronization inevitably starts from the primary $q\bar{q}$-pair in SIA, leading to equal
rates for oppositely charged hadrons, say, e.g.~$\pi^+$ and $\pi^-$ mesons.
This symmetry is broken in SIDIS which proceeds at LO through a single (anti)quark line, 
$\gamma^* q\to q$, and, hence, favors the production of a $\pi^+$ compared to a $\pi^-$ meson
on a proton target. Depending on the choice of the target $N$ in SIDIS, the associated PDFs 
can be viewed as individual weights for different quark and antiquark flavors, albeit with
uncertainties as PDFs are also extracted from data.
In addition, SIDIS data also probe the FFs at complementary energy scales, see Sec.~\ref{chap3:data}.

The neutral- and charged-current SIDIS cross section in terms of structure functions is given by 
\cite{deFlorian:2012wk,Bonino:2025qta, Goyal:2026ccx}
\begin{eqnarray}
\label{eq:sidis-xsec}
\frac{d\sigma^h}{dx\, dy\, dz} =
\frac{2\pi\alpha^2}{Q^2} \eta^{\mathrm{NC/CC}}
\left[ \frac{1+(1-y)^2}{y} 2F_1^{h,\mathrm{NC/CC}}(x,z,Q^2) 
+ \frac{2 (1-y)}{y} F_L^{h,\mathrm{NC/CC}}(x,z,Q^2) + l_\ell\frac{1-(1-y)^2}{y} F_3^{h,\mathrm{NC/CC}}(x,z,Q^2)\right],
\end{eqnarray}
with $x$ and $y$ denoting the usual DIS scaling variables $(Q^2=S x y)$, and
where \cite{Altarelli:1979kv,Nason:1993xx, Furmanski:1980cm} the momentum fraction $z$  
is now defined as $z\equiv p_h\cdot p_N/p_N\cdot q$
in terms of the four-momenta of the incoming nucleon and the outgoing hadron, 
$p_N$ and $p_h$, respectively. 
In the case of CC (SI)DIS with an outgoing neutrino, the kinematic variables can still be reconstructed experimentally 
using, e.g., the Jacquet-Blondel method \cite{Amaldi:1979yh}.
$l_\ell=\pm 1$ corresponds to the lepton number, while the factors 
$\eta^{\mathrm{NC/CC}}$ related to the gauge boson propagators can be found in \cite{ParticleDataGroup:2024cfk}.

The NC structure functions in Eq.~(\ref{eq:sidis-xsec}) can be decomposed
further into contributions from pure photon or $Z$ boson exchange, as well as their 
interference, as
\begin{eqnarray}
\label{eq:f1sidis_NC}
 F_{1,L}^{h,\mathrm{NC}} &\!\!=\!\!& F_{1,L}^{h,\gamma}-\left(g_V^\ell-l_\ell\lambda_\ell g_A^\ell\right)\,\eta_{\gamma Z}\, F_{1,L}^{h,\gamma Z} + \left(g_V^{\ell\,2}+g_A^{\ell\,2}-2 l_\ell \lambda_\ell g_V^\ell g_A^\ell\right)\,\eta_{Z}\, F_{1,L}^{h,Z},  \\
\label{eq:f3sidis_NC}
F_{3}^{h,\mathrm{NC}} &\!\!=\!\!& -\left(g_A^\ell-l_\ell\lambda_\ell g_V^\ell\right)\,\eta_{\gamma Z}\, F_{3}^{h,\gamma Z}+\left[2 g_V^\ell g_A^\ell - l_\ell \lambda_\ell \left(g_V^{\ell\,2}+g_A^{\ell\,2}\right)\right]\,\eta_{Z}\, F_{3}^{h,Z},
\end {eqnarray}
where $g_{V}^\ell$ and $g_{A}^\ell$ denote the lepton's vector and axial coupling to
the $Z$ boson, respectively, which again can be found in \cite{ParticleDataGroup:2024cfk}. 
$\lambda_\ell=\pm 1$ indicates the helicity of the incoming lepton $\ell$.
Similarly, the CC structure functions can be written in terms of the contribution 
from $W^{\pm}$ boson exchange as
\begin{equation}
F^{h,\mathrm{CC}}_{1,L} = F^{h,W^\pm}_{1,L}\,\,\text{and}\,\,\, F^{h,\mathrm{CC}}_3 = F^{h,W^\pm}_3 .
\label{eq:f1sidis_CC}
\end{equation}

In all cases, the structure functions in Eqs.~(\ref{eq:f1sidis_NC})-(\ref{eq:f1sidis_CC}) 
can be factorized in terms of the PDF $f_a^N$ for a parton $a$ inside the nucleon $N$, 
the hard scattering coefficient functions $\mathbb{C}_{a b}^{i,k}$ describing the 
calculable parton-level process $a\rightarrow b X$ as a series expansion in $\alpha_s$,
and the FF $D_{b}^h$ for a parton $b$ to produce the identified hadron $h$, $D_{b}^h$.
Schematically, this double-convolution reads
\begin{equation}
F_i^{h,k}(x,z,Q^2) = \!\sum_{a,b= q,\bar{q},g}\!
\,f_a^N\otimes \mathbb{C}_{a b}^{i,k}\otimes D_{b}^h\,\,\text{where}\,\,
i=1,L,3\,\,\text{and}\,\,k=\mathrm{NC},\mathrm{CC}\,.
\label{eq:factorization_SIDIS}
\end{equation}
Once again, the scales $\mu_f$ and $\mu_r$, for brevity suppressed in Eq.~(\ref{eq:factorization_SIDIS}), should be chosen to be of the order of the
only available hard scale $Q$ in SIDIS.
It is worth mentioning that, similarly to SIA, the contributions involving a 
fragmenting gluon only appear at NLO accuracy and beyond, thus reducing the 
sensitivity of SIDIS data to $D^h_g$.
As we have already stressed, the SIDIS structure functions $F_i^{h,k}$ in Eq.~\ref{eq:factorization_SIDIS} 
involve different combinations of PDFs and FFs depending on the target $N$ and the choice of NC or CC, 
which can potentially improve our knowledge of the flavor and quark/antiquark dependence of FFs.
While current SIDIS data, see Chap.~\ref{chap3:data}, were collected at comparatively low, 
fixed-target energies $\sqrt{S}$ and are only sensitive to photon-mediated NC 
contributions in (\ref{eq:sidis-xsec}), future data from an Electron-Ion Collider (EIC) 
\cite{AbdulKhalek:2021gbh}, currently under construction at Brookhaven National Laboratory (BNL), 
will cover the full wealth of NC and CC SIDIS as outlined above.
The potential impact of future CC SIDIS data in the determination of FFs has been noted 
in \cite{Bonino:2025tnf}.

The expansion in the strong coupling of the perturbatively calculable 
coefficient functions $\mathbb{C}_{ab}^{i,k}$ in Eq.~(\ref{eq:factorization_SIDIS})
reads
\begin{equation}
\mathbb{C}_{ab}^{i,k}
=
\mathbb{C}_{ab}^{i,k,(0)}
+
\frac{\alpha_s(\mu_r)}{2\pi}\,
\mathbb{C}_{ab}^{i,k,(1)}
+
\left(
\frac{\alpha_s(\mu_r)}{2\pi}
\right)^2
\mathbb{C}_{ab}^{i,k,(2)}
+
\mathcal{O}(\alpha_s^3)\,.
\label{eq:sidis-coeff-expansion}
\end{equation}
The $\mathcal{O}(\alpha_s)$ NLO corrections were obtained in 
\cite{Altarelli:1979kv, Baier:1979sp, deFlorian:1997zj,deFlorian:2012wk},
for which also analytical expressions in Mellin moment space 
can be given, see \cite{deFlorian:1997zj,Anderle:2012rq}. The latter now involve a double
Mellin transform, one for each kinematic variable $x$ and $z$.
The NNLO coefficient functions $\mathbb{C}_{ab}^{i,k,(2)}$
for photon-mediated SIDIS as well as for the electroweak contributions 
were obtained in \cite{Goyal:2023zdi, Bonino:2024qbh, Goyal:2024emo,Bonino:2025qta, Goyal:2026ccx}.
Numerical implementations for fully-differential SIDIS were obtained 
recently in \cite{Gao:2026tnd, Dong:2026uvw}. 

In experiment, it is common to measure the hadronic multiplicity $d^3M^h/dz\,dx\,dQ^2$, 
defined as the ratio of the SIDIS cross section (\ref{eq:sidis-xsec}) in a certain bin 
of $Q^2$, $x$, and $z$ to the fully inclusive DIS rate,
\begin{equation}
\label{eq:sidis-mult}
\frac{d^3M^h}{dz\,dx\,dQ^2}=\frac{d^3\sigma^h/dz\,dx\,dQ^2}{d\sigma/dx\, dQ^2}\,,
\end{equation}
rather than quoting the cross section itself, see Chap.~\ref{chap3:data}.
This has the advantage that certain uncertainties at least partially cancel in
the ratio (\ref{eq:sidis-mult}).
Strictly speaking, Eq.\ (\ref{eq:sidis-xsec}) and the
factorized framework adopted in Eq.\ (\ref{eq:factorization_SIDIS})
only apply to hadron production in the current fragmentation region, i.e.,
the hadron should emerge close in rapidity to the parent parton.
This is usually ensured by a cut $x_F>0$ on the Feynman-variable representing 
the fractional longitudinal momentum of the produced hadron
in the c.m.s.\ frame of the nucleon and the virtual boson.
For a more detailed account in the context of TMD factorization and using
the hadron's rapidity, see \cite{Boglione:2016bph}.
If desired, factorization can be extended to the
target fragmentation regime as well by transforming to the 
variable $z \rightarrow z \equiv \frac{E_h}{E_N (1-x)}$,
\cite{Graudenz:1994dq, Trentadue:1993ka, deFlorian:1995fd}
but at the expense of introducing a new set of non-perturbative 
fracture functions \cite{Trentadue:1993ka}.
They can be viewed as describing the conditional probability of finding a 
specific parton inside the incoming nucleon $N$ while simultaneously 
detecting a hadron $h$ in the final state, i.e., some kind of hybrid 
between PDFs and FFs.
Fracture functions obey an inhomogeneous scale evolution but otherwise
very little is known about them.

The resummation of large threshold logarithms in the coefficient functions 
$\mathbb{C}_{ab}^{i,k}$ has been performed in 
\cite{Sterman:2006hu,Anderle:2012rq,Abele:2021nyo,Goyal:2025bzf,Forte:2026tva}
and was pushed to N$^3$LL accuracy in \cite{Abele:2022wuy}. 
These results have been used to derive useful approximate NNLO and N$^3$LO corrections
for SIDIS by re-expanding the resummed expressions \cite{Abele:2022wuy,Forte:2026tva}.
In contrast to the full NNLO corrections, analytical Mellin moments are available in
case of the approximate higher-order results.
A numerical analysis based on the NNLL expressions matched to the 
exact FO NNLO results has demonstrated very good perturbative stability 
and a reduced scale dependence for SIDIS \cite{Goyal:2025bzf}.
To the best of our knowledge, all-order resummations to SIDIS in the limit of small
momentum fractions $x,z\to 0$ have not been considered so far.

We note that the coefficient functions for SIDIS are also available in case
of a longitudinally polarized nucleon $N$ and identified hadron $h$.
Results at NLO and NNLO accuracy can be found in 
\cite{deFlorian:1997zj,deFlorian:2012wk} and
\cite{Bonino:2024wgg,Goyal:2024emo,Bonino:2025bqa,Goyal:2026ccx}, respectively,
and corresponding threshold resummations have been addressed in \cite{Anderle:2013lka}
up to NLL accuracy.

As in SIA, factorization in SIDIS only holds if the hard scale $Q$ is large enough,
say, at least of the order of $1\,\mathrm{GeV}$ or more to keep potential power 
corrections at bay. 
Compared to SIA, there are now two sources of kinematical $1/Q^2$ corrections
stemming from both the so far neglected mass $m_N$ and $m_h$ of the target
nucleon and the identified hadron, respectively. The former have been known for a long time
in DIS as target mass corrections \cite{Georgi:1976ve} that mainly affect 
fixed-target data at comparatively low $\sqrt{S}$ as $x\to 1$. 
Power corrections due to the finite $m_h$ have been considered, for instance, in 
\cite{Albino:2006wz,Albino:2008gy,Accardi:2009md}.
In addition, there will be genuine power corrections to SIDIS stemming from higher-twist 
multi-parton correlations.

\subsubsection{Hadron production at high transverse momentum in hadronic collisions}
\label{sec:pp-theory}
\begin{figure}[ht!]
\centering
\includegraphics[width=.3\textwidth]{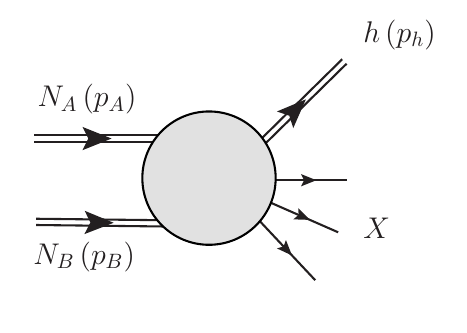}
\caption{Schematic representation of high-$p_T$ hadroproduction,
$N_A N_B\rightarrow h X$,
where the grey blob represents all Feynman diagrams contributing to the production of 
an identified hadron $h$ at a given order in pQCD.}
\label{fig:SIH}
\end{figure}

%
The single-inclusive production of a hadron $h$ at
high transverse momentum $p_T$ in hadron-hadron collisions, $N_A N_B\rightarrow h X$,
depicted in Fig.~\ref{fig:SIH}, can also be described within the framework of pQCD. 
It plays a very crucial role in the extraction of FFs. As opposed to SIA and SIDIS, hadroproduction
has the advantage of involving contributions with a fragmenting gluon already at LO accuracy, 
and can even be dominated by the partonic process $g g \rightarrow g X$ at low and intermediate 
values of $p_T$ \cite{Sassot:2009sh,Sassot:2010bh} where the bulk of the available data is located,
see Chap.~\ref{chap3:data}.
This additional sensitivity allows one to determine the gluon FF, which is otherwise only indirectly
and, hence, very weakly constrained through the scale evolution equations (\ref{eq:singletevol}) for FFs 
and sub-leading QCD corrections in SIA and SIDIS.
%
%

In a similar fashion to the processes discussed previously, and up to corrections suppressed 
by inverse powers of the hard scale $p_T$, the relevant differential 
hadroproduction cross section can once again be cast into 
factorized form as \cite{Aversa:1988vb,Jager:2002xm, deFlorian:2002az}
\begin{equation}
\label{eq:pp-xsec}
E_h \frac{d^3 \sigma}{dp_h^3}  =\frac{1}{2\pi p_T}\frac{d^2\sigma}{dp_T dy}
= \sum_{a,b,c}\int\frac{dz}{z^2}\int dx_a\int dx_b\, f_a^{N_{A}}(x_a,\mu_f) \, f_b^{N_{B}}(x_b,\mu_f)\, 
d\hat{\sigma}_{ab}^{c}(x_a,x_b,z,\mu_r,\mu_f) \,D_c^{h}(z,\mu_f)\,,
\end{equation}
where $p_h$ denotes the spatial momentum of the detected hadron and $y$ its rapidity, which
coincides with the more commonly used pseudo-rapidity $\eta$ in the limit $m_h\to 0$.
$p_T$ sets the hard scale and is defined as the component of $p_h$ that is perpendicular 
to the direction of the collision axis of the two colliding beam particles or to the 
direction of the incoming beam in case of a fixed-target experiment.
It should be noted that convolutions in Eq.~(\ref{eq:pp-xsec}) now involve two sets of PDFs in addition 
to the FF $D_c^{h}$, $f_a^{N_{A}}$ and $f_b^{N_{B}}$, corresponding to partons $a$ and $b$ inside the 
incoming nucleons $N_A$ and $N_B$, respectively. 

The sum in Eq.~(\ref{eq:pp-xsec}) runs over all contributing partonic channels $a+b\to c + X$, with 
$d\hat{\sigma}_{ab}^{c}$ the associated partonic cross section, which can be expanded 
as a power series in the strong coupling $\alpha_s$. 
Due to the more intricate nature of the process, involving three non-perturbative quantities 
and several partonic channels opening up already at the LO level, the hadroproduction 
cross section was until recently only known up to NLO (${\cal{O}}(\alpha_s^3)$) accuracy
\cite{Aversa:1988vb, Jager:2002xm, deFlorian:2002az}. For similar reasons, analytic expressions for the
cross section in Mellin space, which would involve Mellin-transforming the variables $x_a,\,x_b$ and 
$z$, are not known. Therefore, one has to resort to pre-calculated grids in fits of FFs
as was mentioned in Sec.~\ref{subsec:solving-rge}. This is, however, also the case if one 
prefers to work in momentum space since the numerical evaluation of Eq.~(\ref{eq:pp-xsec})
is rather time-consuming due to the multiple integrations and the complexity of the expressions.
Recently, the first complete NNLO calculation for single-inclusive hadroproduction 
was performed by numerical means in \cite{Czakon:2025yti} but still awaits an efficient
implementation into the global analysis framework for FFs.

Enhanced threshold logarithms in single-inclusive hadroproduction arise when the invariant mass squared
of the system recoiling against the fragmenting parton, $\hat{s}_X=(p_a+p_b-p_c)^2$, 
becomes small in the process $a+b\to c + X$.
Their all-order resummation up to NLL accuracy was first implemented in 
\cite{deFlorian:2005yj, Catani:2013vaa} and subsequently extended to NNLL in \cite{Hinderer:2018nkb}. 
Re-expanded expressions have served as useful surrogates for the full NNLO corrections in the past.

%
%
\begin{figure}[th!]
\centering
\includegraphics[width=.75\textwidth]{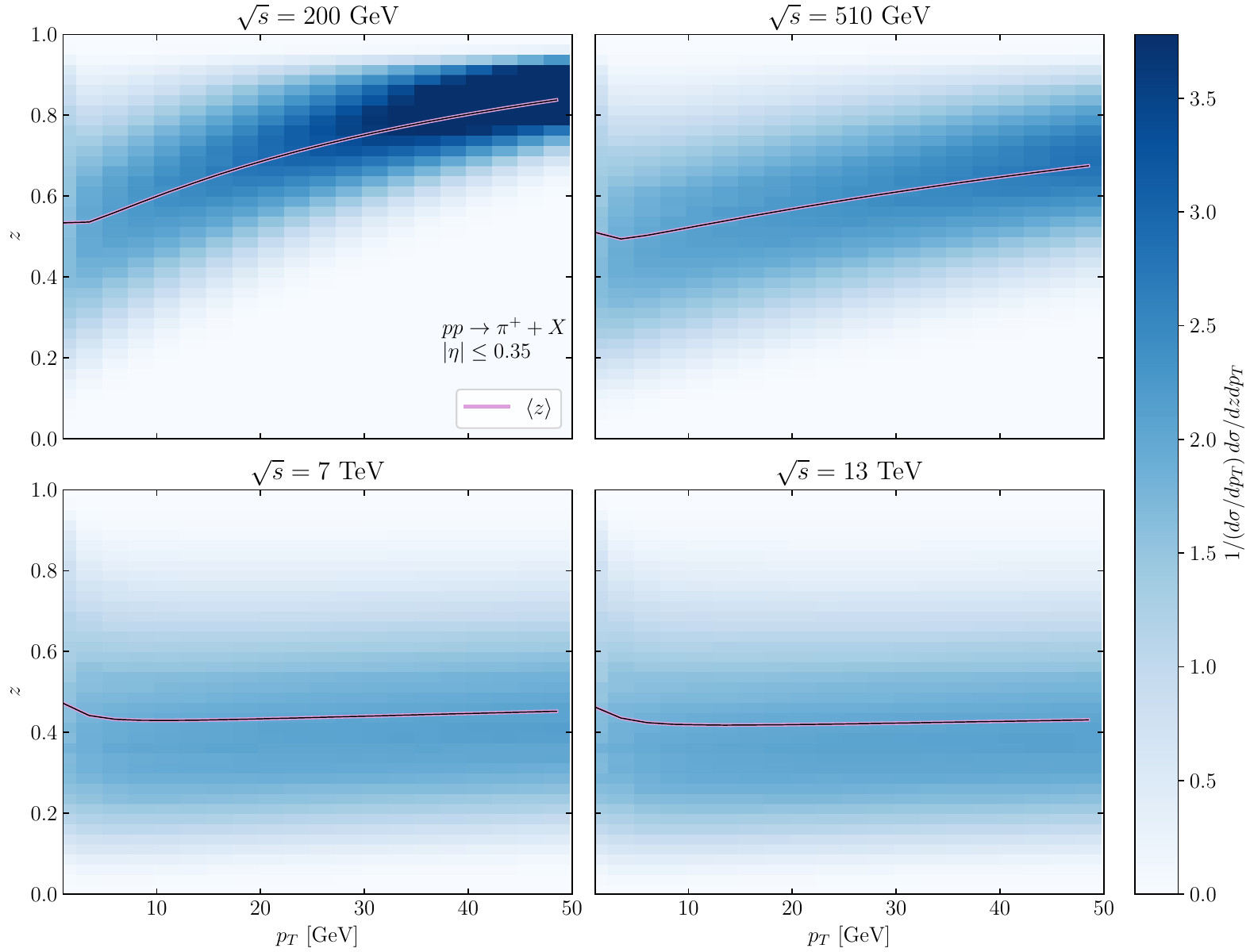}
\caption{The LO cross section for the process $pp\rightarrow\pi^+X$ at mid pseudo-rapidity 
($|\eta\leq0.35|$) as a function of $z$ and $p_T$ for four different values of the 
c.m.s.~energy $\sqrt{S}$ relevant in global fits and normalized to
the $p_T$-differential cross section. The resulting mean value $\langle z \rangle$ for the momentum fraction $z$ as a function of $p_T$ is indicated by the solid lines.}
\label{fig:pp_zdist}
\end{figure}
It is worth recalling that in both SIA and SIDIS at the LO approximation
the FFs are probed locally at a single value of $z$ which is fully determined from measurable quantities.
This is in stark contrast to the situation in hadroproduction, where the FFs $D_c^{h}(z,\mu_f)$ in
(\ref{eq:pp-xsec}) are always evaluated within a convolution, i.e., an integration over a broad range 
of $z$, where only its lower limit $z_{\min}=x_T\cosh\eta $ is determined kinematically,
with $x_T\equiv 2p_T/\sqrt{S}$ the hadron's scaled transverse momentum.
To better understand the seemingly reduced sensitivity of hadroproduction data
to FFs, we display in Fig.~\ref{fig:pp_zdist} LO estimates of the cross section (\ref{eq:pp-xsec}) 
for $pp\rightarrow \pi^+X$ at mid pseudo-rapidity and differential in $p_T$ and $z$.
In each bin of $p_T$ and $z$ the results are normalized to the respective $z$-integrated 
cross section $d\sigma/dp_T$ in that bin. The darker the shading the larger the contribution
of a certain bin in $z$ for a given $p_T$. The solid curves show the resulting
mean value $\langle z \rangle$ as a function of $p_T$ in each of 
the four panels that are obtained for different values of the c.m.s.~energy $\sqrt{S}$
pertinent to existing data used in global fits, see Chap.~\ref{chap3:data} and \ref{chap4:fits}.
It is interesting and, perhaps, at first sight also surprising to note that 
the FFs are predominantly probed in a rather narrow range around fairly large 
values of $\langle z \rangle$ well above their lower kinematic limits $z_{\min}=x_T\cosh\eta$. 
While the $\langle z \rangle$ distribution is rather flat at the highest energies $\sqrt{S}$
in the shown range of $p_T$, it increases towards its upper limit $z_{\max}=1$ in the two
upper panels of Fig.~\ref{fig:pp_zdist} as $x_T$ approaches unity.
The results obtained in Fig.~\ref{fig:pp_zdist} can be largely understood from the
interplay of the FFs with the behavior of the PDFs entering (\ref{eq:pp-xsec}).
The smaller $x_T$ the more dominant is the contribution of gluon-gluon scattering
to (\ref{eq:pp-xsec}) driven by the strong rise of the gluon density towards small 
values of $x_{a,b}$. On average this leads to fairly small values 
of the available partonic c.m.s.\ energy in the hard scattering process which
in turn demands rather large momentum fractions $z$ to produce the observed
hadron $h$ at high-$p_T$.

From a theoretical perspective, it is good news that the region $z\to 0$ 
contributes very little to the hadroproduction cross section irrespective of $\sqrt{S}$
as this avoids any worries about potentially large logarithmic contributions to
the scale evolution of FFs discussed in Sec.~\ref{subsec:resum}.
Hence, it should be possible to obtain reliable estimates for (\ref{eq:pp-xsec}) in
FO pQCD as long as one stays within the limits of the factorization theorem by placing
a sufficiently large cut on the minimum $p_T$ of the analyzed data. This should in turn
also avoid sizable power corrections to (\ref{eq:pp-xsec}), such
as those stemming from the finite hadron mass $m_h$ which were considered in 
\cite{Albino:2008fy,Albino:2008gy} in the context of fits of FFs for various hadron species.

Finally, we note that one proposed way to delineate the leading-twist framework
based on factorization as outlined above from other, power-suppressed direct
production mechanisms for the observed hadron $h$
is to study the $x_T$-scaling of the available data, see, e.g.~\cite{Sivers:1975dg,Arleo:2009ch}.
More specifically, the underlying idea is based on the fact that 
in the naive parton model the hard scattering is scale-invariant and 
a dimensional analysis then dictates that the cross section in (\ref{eq:pp-xsec})
obeys a power-law $F(x_T,\eta)/p_T^n$ at fixed $x_T$ and $\eta$ with some function $F(x_T,\eta)$
and a scaling exponent $n$ that is exactly $4$ \cite{Brodsky:1973kr,Sivers:1975dg}.
Thus, data for different $\sqrt{S}$ should all fall on a universal curve when plotted
as $p_T^4\,E_h \frac{d^3 \sigma}{dp_h^3}$ versus $x_T$.
In the real world, the exact scale invariance is broken by QCD corrections, i.e.,
the running of $\alpha_s(\mu_r)$ and the scale evolution of PDFs and FFs.
In FO pQCD this leads to a scaling exponent $n$ that exhibits some nontrivial dependence on $p_T$ and
increases with $x_T$ in a range between 5 and 6 \cite{Arleo:2009ch,Sassot:2010bh}. On the other hand,
power-suppressed contributions to (\ref{eq:pp-xsec}) have estimated exponents of at least $n=6$ and 
are expected to become significant at large $x_T$ and 
not too large values of $p_T$ \cite{Berger:1980qg,Arleo:2009ch}.
While there are some indications for the presence of such power-suppressed 
contributions mainly from low-energy data \cite{Arleo:2009ch}, the overall picture is still inconclusive as
the FO results for $n$ at large $x_T$ can be significantly altered by all-order resummations 
which have not been included in available studies yet.

\subsubsection{Less inclusive hadron production processes}
\label{sec:injet}
%
Here, we briefly review some interesting probes of FFs that opened up more recently both
in terms of their underlying theoretical framework and their first experimental results. 
So far, these processes have seen only occasional, if any use in global analyses but we expect them
to play a more important role in the future.
\paragraph{Hadrons in jets}
As was mentioned in the previous section, one of the main drawbacks of high-$p_T$ inclusive hadron
production is that the LO cross section already involves a convolution over the momentum fraction
$z$, thus obscuring its sensitivity to FFs. 
In this regard, in-jet fragmentation observables offer complementary and more direct constraints.
In this case, one considers the production of an identified hadron $h$ inside a reconstructed jet, 
and studies the cross section differential in $z_h=p_T/p_T^{\mathrm{jet}}$, with 
$p_T$ ($p_T^{\mathrm{jet}}$) denoting the transverse momentum of the hadron (jet). 
The main advantage of these in-jet observables is that, at LO accuracy, they directly 
probe the $z=z_h$ dependence of FFs.

In the case of hadronic collisions, the cross section for the production of an identified hadron
inside a jet can be cast in a similar way as Eq.~(\ref{eq:pp-xsec}) and reads
\begin{equation}\label{eq:h-in-jet-xsec}
\frac{d\sigma^{(\mathrm{jet}\, h)}}{dp_T^{\rm jet}\, d\eta^{\rm jet}\, dz_h}=
\sum_{a,b,c}
\int\frac{dz_p}{z_p^2}
\int dx_a
\int dx_b\,
f_a^{N_A}(x_a,\mu_f)\,f_b^{N_B}(x_b,\mu_f)\,
d\hat{\sigma}^{(\mathrm{jet\,c})}_{ab}
(x_a,x_b,z_p,\mu_r,\mu_f, R)\,
D_c^h\left(\frac{z_h}{z_p},\mu_f\right)\,,
\end{equation}
with the main difference being that the partonic cross section
$d\hat{\sigma}^{(\mathrm{jet\,c})}_{ab}$ now corresponds to the production of a final-state jet 
containing a parton $c$ with transverse momentum $p_T^c=z_p p_T^{\mathrm{jet}}$, i.e.,
$z_p$ is the partonic analog of $z_h$. Since the partonic cross section involves the reconstruction of
a jet in the final state, it depends on the algorithm adopted to define the jet and therefore on a
generic size parameter $R$. At LO accuracy, the jet is formed by a single parton, which is also the
one fragmenting into the identified hadron $h$, thus fixing $z_p=1$ and hence the argument of the FFs to
$z=z_h$.

The inclusive production of identified hadrons inside a fully reconstructed jet has been studied in
Refs.~\cite{Arleo:2013tya, Kaufmann:2015hma, Kang:2016ehg, Czakon:2025yti}. In \cite{Arleo:2013tya}, the NLO cross section was 
obtained numerically using a Monte Carlo phase-space integrator to evaluate Eq.~(\ref{eq:h-in-jet-xsec}). A similar approach was used in \cite{Czakon:2025yti}, extending the perturbative accuracy to NNLO.
Instead, analytical expressions were derived in Refs.~\cite{Kaufmann:2015hma, Kang:2016ehg} 
using the approximation that the jet is sufficiently collimated. 
At NLO accuracy, these results can be schematically written in terms of the jet-size parameter $R$ 
as $A + B \ln R + \mathcal{O}(R^2)$ with some coefficients $A$ and $B$. 
If the jet is sufficiently narrow, i.e., $R \ll 1$, power corrections of $\mathcal{O}(R^2)$ 
can be neglected. 
Within this narrow-jet approximation, the cross section for the process
$N_AN_B \to (\mathrm{jet}\, h)X$ may be written as
\begin{equation}\label{eq:h-in-jet-xsec-NJA}
\frac{d\sigma^{(\mathrm{jet}\, h)}}{dp_T^{\rm jet}\, d\eta^{\rm jet}\, dz_h}=
\sum_{a,b,c}
\int\frac{dz_c}{z_c^2}
\int dx_a
\int dx_b\,
f_a^{N_A}(x_a,\mu_f)\,f_b^{N_B}(x_b,\mu_f)\,
d\hat{\sigma}^{c}_{ab}
(x_a,x_b,z_c,\mu_r,\mu_f)\,
\mathcal{G}_c^h(z_c,z_h,\mu_f,R)\,,
\end{equation}
with $d\hat{\sigma}^{\,c}_{ab}$ corresponding to the same partonic cross section for the 
process $a+b\to c+X$ appearing in Eq.~(\ref{eq:pp-xsec}). Note that the FF for parton $c$ 
in Eq.~(\ref{eq:h-in-jet-xsec}) is now replaced by the fragmenting-jet function $\mathcal{G}_c^h$,
which codifies the transition from parton $c$ to the final-state jet and the identified hadron inside
that jet. The fragmenting-jet function naturally inherits the dependence on the jet algorithm and the
parameter $R$. In this case, the convolution variable $z_c$ corresponds to the fraction of the
transverse momentum of parton $c$ carried by the reconstructed jet, that is, 
$p_T^{\mathrm{jet}}=z_c p_T^c$. 
To NLO accuracy, $\mathcal{G}_c^h$ in Eq.~(\ref{eq:h-in-jet-xsec-NJA}) can be decomposed in 
terms of standard FFs as
\begin{equation}\label{eq:h-in-jet-G}
\mathcal{G}_c^h(z_c,z_h,\mu_f,R)
=
\sum_e
j_{c\to e}(z_c,R,\mu_f)
\sum_{c'}
\int_{z_h}^{1}
\frac{d\xi}{\xi}\,
\tilde{j}_{e\to c'}(\xi,R,\mu_f)\,
D_{c'}^h
\left(
\frac{z_h}{\xi},
\mu_f
\right).
\end{equation}
Schematically, Eq.~(\ref{eq:h-in-jet-G}) can be interpreted as the sequential formation of a jet from the scattered parton $c$
followed by the hadronization of one of the partons inside the reconstructed jet. The jet functions $j_{c\rightarrow e}$ 
and $\tilde{j}_{e\rightarrow c'}$ describe, respectively, the formation of the jet consisting of a parton $e$ carrying a fraction
$z_c$ of the momentum of the parent parton and the fragmentation of parton $e$ into parton $c'$ inside the jet. 
$D^h_{c'}$ denotes the usual FF into an identified hadron $h$. Both sets of jet functions can be calculated in pQCD, 
and their corresponding expressions may be found in Ref.~\cite{Kaufmann:2015hma}.

Besides hadronic collisions, hadron-in-jet production has also been studied for $e^+e^-$ collisions within SCET
\cite{Procura:2009vm, Jain:2011xz, Procura:2011aq}. In that context, the cross section can be once again written
in terms of fragmenting-jet functions, which can be matched to the standard FFs.
More recently, the cross section for hadron-in-jets production both in the case of $e^+e^-\rightarrow 2$ jets 
and $e^+e^-\rightarrow 3$ jets has been calculated numerically up to NNLO accuracy in pQCD \cite{Bonino:2026dvr}.
\paragraph{Hadron plus vector boson production in hadronic collisions}
Collinear factorization for the associated production of an identified hadron and a vector boson $V$ 
in hadronic collisions proceeds along the same lines as in Eq.~(\ref{eq:pp-xsec}) 
apart from replacing the partonic cross sections by $d\hat{\sigma}^{a+b\to c+V+X}$. 
Unlike inclusive hadron production, requiring an additional gauge boson in the final state allows one 
to reconstruct the momentum fraction $z$ at the Born level in terms of experimentally accessible 
quantities \cite{Ochoa-Oregon:2023ktx}. Here, $V$ and the outgoing parton $c$, 
that fragments into the observed hadron $h$, are back-to-back in transverse momentum, 
resulting in $z= p_T^h/p_T^V$. 

Photon+hadron and di-hadron production were calculated at NLO QCD accuracy 
in \cite{Binoth:2002wa}, comprising the direct-photon and parton-to-photon fragmentation
\cite{Gluck:1992zx,Bourhis:1997yu} components, respectively. More recent photon+hadron studies 
have supplemented these predictions with leading QED corrections \cite{Ochoa-Oregon:2023ktx}, while the NNLO QCD predictions for photon-associated hadron production  were obtained in \cite{Fiore:2026pot}.
Fully differential NLO QCD results for $W^{\pm}/Z$+hadron production have subsequently been obtained 
and applied to existing $Z$ boson plus charged-hadron-in-jet and $W^{\pm}$ boson plus charmed meson 
measurements \cite{Caletti:2024xaw}. 
The associated production of a $W^{\pm}$ boson and a charmed meson, which is also of particular interest
for determining the strangeness PDF, was recently calculated up to NNLO accuracy in \cite{Generet:2025bqx}.
\paragraph{Di-hadron production}
The production of two identified hadrons provides information on collinear FFs complementary
to that obtained in single-inclusive measurements. In $e^+e^-$ annihilation, and assuming 
the two identified hadrons originate from separately resolved partons, the
cross section takes the schematic factorized form
\begin{equation}\label{eq:dih-ee}
  d\sigma^{e^+e^-\to h_1h_2X}
  =
  \sum_{i,j}
  d\hat{\sigma}^{e^+e^-\to ijX}
  \otimes D_i^{h_1}\otimes D_j^{h_2}.
\end{equation}
where the convolutions run over the fragmentation momentum fractions corresponding to
$h_1$ and $h_2$. 
At the Born approximation, the virtual photon or $Z$ boson produces
a primary $q\bar{q}$ pair, whose independent fragmentation gives
rise to the two observed, back-to-back hadrons. 
For photon exchange, the flavor structure is proportional to
$\sum_q e_q^2[D_q^{h_1}D_{\bar q}^{h_2} +D_{\bar q}^{h_1}D_q^{h_2}]$.
Hadron-pair measurements therefore probe combinations of quark
and anti-quark FFs complementary to those entering SIA, see Sec.~\ref{sec:sia-theory}.
Some care is needed, however, if both SIA and two-hadron data are used simultaneously
in a fit of FFs as a sizable fraction of events with two (or more) identified hadrons 
certainly also contributes to the  experimental result for single-inclusive hadron yields.
This can lead to pronounced correlations of the data
which need to be taken care of.

Experimental results for (\ref{eq:dih-ee}) are usually given in terms of the 
invariant mass $m_{hh}$ of the di-hadrons rather than binning them in their 
individual transverse momenta.
For light hadrons such as pions and kaons, a large enough $m_{hh}$ automatically ensures
the requirement of independent fragmentation.
The two-hadron cross section (\ref{eq:dih-ee}) has been calculated at NLO accuracy in 
\cite{deFlorian:2003cg}. More recently, its angular dependence for $0<\theta_{hh}<\pi$
was obtained at $\mathcal{O}(\alpha_s^2)$ in~\cite{Ju:2025jxi}.

In hadronic collisions, the factorized cross section can be expressed analogously 
to Eq.~(\ref{eq:dih-ee}) with the inclusion of the additional convolutions due to the 
corresponding PDFs. Implementations for hadron-pair production at NLO accuracy
were first developed in Refs.~\cite{Chiappetta:1996wp, Owens:2001rr,Binoth:2002ym},
and NNLO predictions differential in $m_{hh}$ are now available as well \cite{Czakon:2025yti}.
Relevant threshold logarithms have been resummed at NLL accuracy
with partial extensions to NNLL \cite{Almeida:2009jt,Hinderer:2014qta}.

It is worth stressing again that the factorized expression in Eq.~(\ref{eq:dih-ee}) only holds
if the two hadrons originate from different and well separated partons. 
The case of two nearly collinear hadrons with small invariant mass $m_{hh}$ is also
of considerable phenomenological interest, but theoretically more involved.
Here, the contribution from the double fragmentation of a single parton must 
be considered as well. This kind of hadronization is described by a set of non-perturbative
di-hadron FFs \cite{Bianconi:1999cd}. Hence, if the di-hadron cross section is 
integrated in $m_{hh}$ and the separation of the observed hadrons is not 
ensured otherwise by appropriate angular cuts, both production mechanisms contribute
to the cross section \cite{deFlorian:2003cg}.

\section{Experiments and available data}\label{chap3:data}
In this section we give an overview of the relevant experimental data for the 
processes outlined in Sec.~\ref{sec:processes} and commonly used to determine FFs.
In what follows, we limit ourselves to hadron species that are most abundantly 
produced in experiment and, hence, allow for meaningful extractions of the 
corresponding sets of FFs with uncertainties and tests of their fundamental universality property.
More specifically, we mainly concentrate on pion and kaon production in case 
of identified hadrons. Since many experiments have no means of identifying the hadron, 
also data for unidentified (charged) hadrons play an important role phenomenologically. 
For completeness, we also list available results for $\eta$ mesons, 
(anti)protons, and $\Lambda$ and $\bar{\Lambda}$ baryons, 
where also a fair amount of data and some theoretical analyses exist. 

A summary with selected data sets for the production of these hadron species in
SIA, SIDIS, and $pp$/$p\bar{p}$ collisions is presented 
in Tab.~\ref{tab:collinear-ff-data}.
We note that almost all data sets related to high-energy physics experiments 
can be conveniently retrieved from the \texttt{HEPData} repository.
A useful and comprehensive compilation of all available SIA data, albeit slightly
outdated, can be found in \cite{Lafferty:1995jt}. 
For each data set, the two rightmost columns in Tab.~\ref{tab:collinear-ff-data} 
show the range in momentum fraction $z_h$, estimated at LO accuracy,
and the typical scale $\mu_f$ probed in fits of the corresponding FFs.
In case of hadronic collisions, no values for $z_h$ are specified
for single-inclusive hadron production data as the momentum fraction is not directly
accessible experimentally. But we refer the reader to Sec.~\ref{sec:pp-theory},
in particular, Fig.~\ref{fig:pp_zdist}, for estimates of the relevant mean $\langle z_h\rangle$.
The values for $\mu_f$ in Tab.~\ref{tab:collinear-ff-data} are taken as $Q$ for SIA and SIDIS, 
$p_T^h$ for inclusive hadroproduction, and $p_T^{\rm jet}$ for hadron-in-jet data. 
They refer to the typical choice for $\mu_f$ made in most fits of FFs 
but it should be kept in mind that $\mu_f$ is in principle completely arbitrary. 
Variations around these default values are often used as an estimator of 
missing higher-order corrections and constitute an important part of the
theoretical uncertainties associated with computations of cross sections within pQCD.

\begin{figure}[ht!]
\centering
\includegraphics[width=.8\textwidth]{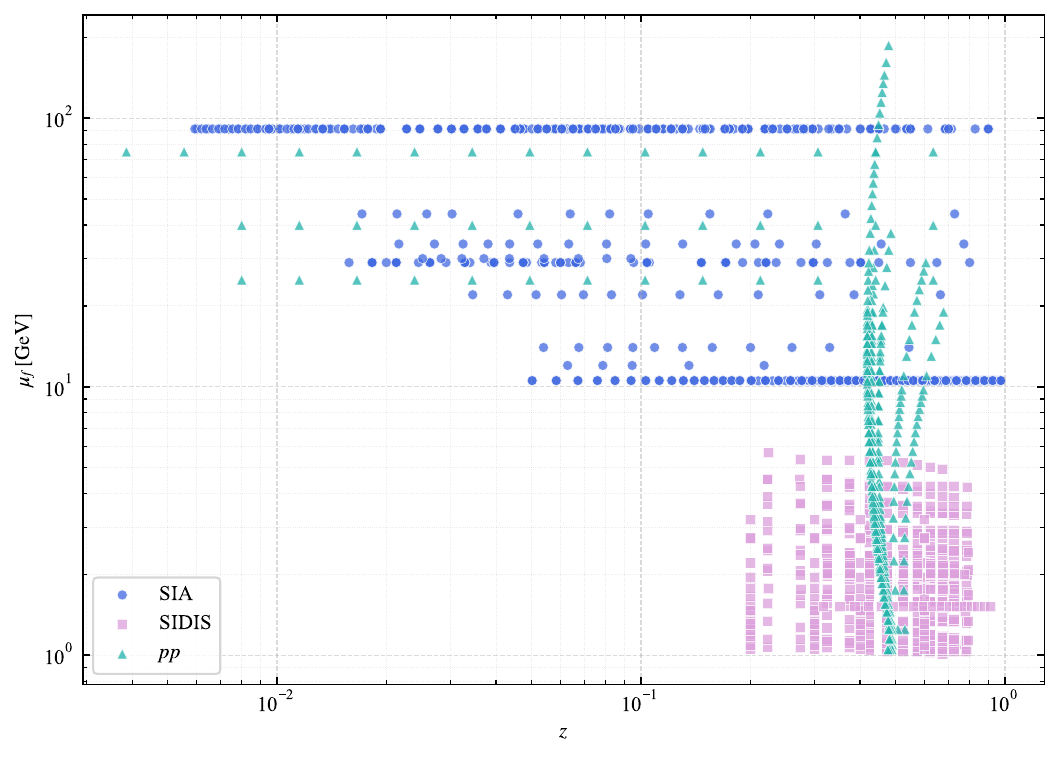}
\caption{Visualization of the kinematic coverage of the currently available 
data relevant for the extraction of parton-to-pion FF, 
see Tab.~\ref{tab:collinear-ff-data}, in the $z$-$\mu_f$ plane.
The $z$-values for single-inclusive hadroproduction refer to the
mean $\langle z\rangle$ estimated at LO accuracy, see Fig.~\ref{fig:pp_zdist} and text.}
\label{fig:FFs_coverage}
\end{figure}

The kinematical coverage of the different processes and data sets 
in the $z$-$\mu_f$ plane is also illustrated in Fig.~\ref{fig:FFs_coverage}
in case of pion production. Only data corresponding to a hard scale of at least
$1\,\mathrm{GeV}$ are shown. 
The complementarity of pion-in-jet and single-inclusive
pion production data is clearly visible. While the latter span a large range
in $\mu_f$ but mainly probe the FFs locally at a mean $\langle z\rangle\approx 0.5$ and
irrespective of $\sqrt{S}$, the in-jet data provide a broad scan in $z$ for various fixed
values of $\mu_f$ similar to SIA. SIDIS data currently contribute only 
below $\mu_f\simeq 10\,\mathrm{GeV}$ as all results stem from fixed-target experiments 
at comparatively low values of $\sqrt{S}$. 
Here, we expect the EIC \cite{AbdulKhalek:2022laj} to significantly expand the coverage 
to much larger scales and to the full electroweak sector of NC and CC contributions, 
see Sec.~\ref{sec:sidis-theory}, in the future.

%
%
\begingroup
\newcolumntype{F}[1]{>{\RaggedRight\arraybackslash}p{#1}}
\newcommand{\FFprocessrow}[1]{%
  \addlinespace[2pt]
  \multicolumn{5}{@{}l}{\bfseries #1}\\[-1pt]
  \midrule
}
\newcommand{\FFfacilityrow}[1]{%
  \multicolumn{5}{@{}l}{\bfseries #1}\\[-1pt]
}
\newcommand{\FFNA}{--}

\setlength{\LTpre}{4pt}
\setlength{\LTpost}{5pt}
\setlength{\LTcapwidth}{\linewidth}
\setlength{\tabcolsep}{2pt}
\renewcommand{\arraystretch}{1.08}
\scriptsize

\begin{longtable}{@{}F{0.295\linewidth}F{0.155\linewidth}F{0.225\linewidth}F{0.135\linewidth}F{0.12\linewidth}@{}}
\caption{Selected experimental measurements relevant to collinear FF for pions, kaons,
unidentified (charged) hadrons, $\eta$ mesons, (anti)protons, and $\Lambda$ and 
$\bar{\Lambda}$ baryons. The two rightmost columns give estimates for the
typical range of $z_h$ and scale $\mu_f$ probed in FFs, see text.
All c.m.s.\ energies $\sqrt{S}$ and scales $\mu_f$ are given in units of GeV unless 
TeV is explicitly stated in case of LHC experiments.}
\label{tab:collinear-ff-data}\\
\toprule\toprule
\textbf{Experiment and reference(s)} & \(\boldsymbol{\sqrt{S}}\) & \textbf{Measured hadrons} & \(\boldsymbol{z_h}\) \textbf{range} & \(\boldsymbol{\mu_f}\) \textbf{range}\\
\midrule
\endfirsthead
\multicolumn{5}{r}{\itshape Table~\ref{tab:collinear-ff-data} continued}\\[-2pt]
\toprule
\textbf{Experiment and reference(s)} & \(\boldsymbol{\sqrt{S}}\) & \textbf{Measured hadrons} & \(\boldsymbol{z_h}\) \textbf{range} & \(\boldsymbol{\mu_f}\) \textbf{range}\\
\midrule
\endhead
\midrule
\multicolumn{5}{r}{\itshape Continued on next page}\\
\endfoot
\bottomrule
\endlastfoot

\FFprocessrow{SIA: \(e^+e^-\to h+X\); see Sec.~\ref{sec:sia-theory}}
\FFfacilityrow{SLAC}
TPC/Two-Gamma \cite{TPCTwoGamma:1984eyl,TPCTwoGamma:1988yjh} & 29 & \(\pi^\pm,K^\pm,p/\bar p,\pi^0\) & \(\sim0.01\)-0.9 & 29\\
SLD \cite{SLD:1998coh,SLD:2003ogn} & 91.2 & \(\pi^\pm,K^\pm,p/\bar p,\Lambda/\bar\Lambda\) & \(\sim0.005\)-1 & 91.2\\
BaBar \cite{BaBar:2013yrg} & 10.54 & \(\pi^\pm,K^\pm,p/\bar p\) & \(\sim0.05\)-1 & 10.54\\

\FFfacilityrow{DESY: PETRA and DORIS}
TASSO \cite{TASSO:1980dyh,TASSO:1981uqa, TASSO:1982bkc,TASSO:1984nda,TASSO:1988jma,TASSO:1989jyt} & 12, 14, 22, 30, 34, 44 & \(h^\pm,\pi^\pm,K^\pm,p/\bar p, \pi^0,\Lambda/\bar\Lambda\) & \(\sim0.02\)-1 & 12-44\\

\FFfacilityrow{KEK}
TOPAZ \cite{TOPAZ:1994voc} & 58 & \(\pi^\pm,K^\pm,p/\bar p\) & \(\sim0.07\)-0.3 & 58\\
Belle \cite{Belle:2020pvy,Belle:2024vua,Belle:2017caf} &  10.58 & \(\pi^\pm,K^\pm,p/\bar p,\eta,\Lambda/\bar\Lambda\) & \(\sim0.1\)-1 & 10.58\\

\FFfacilityrow{CERN:LEP}
ALEPH \cite{ALEPH:1994cbg, ALEPH:1992zhm, ALEPH:1994fts,ALEPH:1996pxg, ALEPH:1999udi,ALEPH:2001tfk} & 91.2 & \(\pi^\pm,K^\pm,p/\bar p, \pi^0, \eta,\Lambda/\bar\Lambda\) & \(\sim0.005\)-0.8 & 91.2\\
DELPHI \cite{DELPHI:1998cgx,DELPHI:2000ahn,DELPHI:1993vpj, DELPHI:1995ase} & 91.2,189 & \(\pi^\pm,K^\pm,p/\bar p, \pi^0, \Lambda/\bar\Lambda\) & \(\sim0.01\)-1 & 91.2, 189\\
OPAL \cite{OPAL:1994zan,OPAL:2000dkf,OPAL:2002isf} & 91.2, 210.7 & \(\pi^\pm,K^\pm,p/\bar p, \pi^0, \eta,\Lambda/\bar\Lambda,h^\pm\) & \(\sim0.005\)-0.8 & 91.2, 201.7\\
L3 \cite{L3:1991nwl,L3:1992pbe,L3:1994gkb} & 91.2 & \(\pi^0,\eta, \Lambda/\bar\Lambda, h^\pm\) & \(\sim0.01\)-1 & 91.2\\

\FFfacilityrow{IHEP: BEPCII}
BESIII \cite{BESIII:2022zit,BESIII:2025mbc} & 2.00-3.67 & \(\pi^\pm, K^\pm\) & \(\sim0.1\)-0.9 & 2.00-3.67\\
\midrule
\FFprocessrow{SIDIS: \(\ell N\to\ell' hX\); see Sec.~\ref{sec:sidis-theory}}
\FFfacilityrow{CERN: fixed-target experiments}
EMC \cite{EuropeanMuon:1984rkv,EuropeanMuon:1985nbu,EuropeanMuon:1991sne} & 13.7-22.9 & \(\pi^\pm, K^\pm,p/\bar p,\Lambda/\bar\Lambda, h^\pm\) & \(\sim0.1\)-1.0 & \(\sim1.5\)-10\\
COMPASS \cite{COMPASS:2024gje,COMPASS:2025bfn} & 17.3 & \(\pi^\pm,K^\pm,h^\pm\) & \(\sim0.2\)-0.85 & \(\sim1\)-10\\
\FFfacilityrow{DESY: HERA fixed-target and collider experiments}
HERMES \cite{HERMES:2001ghm,HERMES:2012uyd} & 7.25 & \(\pi^\pm,\pi^0,K^\pm\) & \(\sim0.1\)-0.8 & \(\sim1\)-4\\
H1 \cite{H1:2007ghd,H1:2009lef,H1:1996kfw} & 300-319 & \(\Lambda/\bar\Lambda,h^\pm\) & \(\sim0.02\)-1 & \(\sim10\)-140\\
ZEUS \cite{ZEUS:1999ies,ZEUS:2009rju,ZEUS:2010mrq, ZEUS:2011cdi} & 300-319 & \( \Lambda/\bar\Lambda,h^\pm\) & \(\sim0.02\)-1 & \(\sim10\)-200\\

\FFfacilityrow{Jefferson Lab: Hall C fixed-target experiments}
Hall C 12-GeV pion \cite{Bhatt:2024prq,HallCSIDIS:2025plr} & 4.47-4.57 & \(\pi^\pm\) & $\sim$0.3-0.7 & \(\sim1.4\)-2.35\\
\midrule
\FFprocessrow{\(pp\) data (and legacy \(p\bar p\)): inclusive hadrons and hadron-in-jet; see Secs.~\ref{sec:pp-theory} and \ref{sec:injet}}
\FFfacilityrow{Fermilab: Tevatron}
CDF \cite{CDF:1988evs,CDF:2005ofj, CDF:2009cxa} & 630, 1800, 1960 & \(\Lambda^0,h^\pm\) & \FFNA & \(\sim0.5\)-20\\

\FFfacilityrow{BNL: RHIC}
PHENIX \cite{PHENIX:2007kqm,PHENIX:2010hvs,PHENIX:2012koo,PHENIX:2014axc,PHENIX:2014qwb,PHENIX:2015fxo} & 62.4, 200, 510 & \(\pi^\pm,\pi^0,\eta,h^\pm\) & \FFNA & \(\sim0.4\)-30\\
STAR \cite{STAR:2006dgg, STAR:2006xud, STAR:2006nmo, STAR:2009vxb, STAR:2012ljf, STAR:2013zyt} & 200 & \(\pi^{\pm},K^\pm,\pi^0,\eta,p/\bar p,\Lambda/\bar\Lambda\) & \FFNA & \(\sim0.2\)-20\\

\FFfacilityrow{CERN: LHC inclusive hadron production}
ALICE  \cite{ALICE:2014juv,ALICE:2017ryd, ALICE:2019hno, ALICE:2020jsh, ALICE:2021est, ALICE:2024vgi} & 2.76, 5.02, 7, 8, 13 TeV & \(\pi^\pm,K^\pm,p/\bar p, \pi^0, \eta,\Lambda/\bar\Lambda\) & \FFNA & \(\sim0.1\)-200\\
ATLAS \cite{ATLAS:2010jvh} & 0.9, 2.36, 7, 8, 13 TeV & \(h^\pm\) & \FFNA & \(\sim0.1\)-200\\
CMS \cite{CMS:2011mry,CMS:2012aa} & 0.9, 2.36, 7, 13 TeV & \(h^\pm\) & \FFNA & \(\sim0.1\)-200\\
LHCb \cite{LHCb:2021abm, LHCb:2023iyw} & 5.02, 13 TeV & \(h^\pm, \eta\) & \FFNA & 0.08-10\\

\FFfacilityrow{CERN: LHC hadron-in-jet and jet-fragmentation measurements}
ATLAS inclusive jets \cite{ATLAS:2011myc,ATLAS:2018bvp,ATLAS:2019rqw} & 5.02, 7, 13 TeV & \(h^\pm\) in jet & \(\sim0.02\)-1 & \(\sim25\)-2500\\
ATLAS \(\gamma\)+jet \cite{ATLAS:2019dsv} & 5.02 TeV & \(h^\pm\) in jet & \(\sim0.02\)-1 & \(\sim60\)-140\\
CMS \(\gamma\)+jet \cite{CMS:2018mqn} & 5.02 TeV & \(h^\pm\) in jet & \(\sim0.02\)-1 & \(>30\)\\
ATLAS \(Z\)+h \cite{ATLAS:2020wmg} & 5.02 TeV & \(h^\pm\) & \FFNA & \(\gtrsim30\)\\
CMS \(Z\)+h \cite{CMS:2021otx} & 5.02 TeV & \(h^\pm\) & \FFNA & \(\gtrsim30\)\\
LHCb \(Z\)+jet \cite{LHCb:2019qoc, LHCb:2022rky} & 8, 13 TeV & \(h^\pm,\pi^\pm,K^\pm,p/\bar p\) in jet & \(\sim0.05\)-1 & \(20\)-100\\
\toprule
\end{longtable}
\endgroup
%
\noindent
The next three subsections give a short account of the experimental efforts worldwide
contributing to the results listed in Tab.~\ref{tab:collinear-ff-data}.

\subsection{Identified particles in $e^+e^-$  collisions}
\label{subsec:sia-data}
As already mentioned in Sec.~\ref{sec:sia-theory}, data on identified hadrons produced in 
$e^+e^-$ collisions play a foundational role in the extraction of FFs, providing the 
cleanest access to these distributions. An extensive amount of data sets on SIA is available 
thanks to worldwide efforts at many major laboratories over the past decades, 
covering a broad range in c.m.s.~energies, $3\,\mathrm{GeV}\lesssim \sqrt{S} \lesssim 200\,\mathrm{GeV}$. 
Measurements of SIA have been performed by experiments at SLAC, DESY, KEK, CERN, and IHEP.

At SLAC, identified-hadron spectra were first measured in the 1980s by TPC/Two-Gamma \cite{TPCTwoGamma:1988yjh}, 
HRS, and MARK-II at the single-ring $e^+e^-$ collider PEP at about $\sqrt{S}=29~\mathrm{GeV}$ \cite{Lafferty:1995jt},
followed by the SLD experiment \cite{SLD:2003ogn} at the first and only linear $e^+e^-$ collider 
operating at the $Z$-boson mass. Data taking in SIA ended in 2008 with the conclusion of the
BaBar experiment \cite{BaBar:2013yrg} at the PEP-II $B$-factory, an asymmetric $e^+e^-$ machine 
with $\sqrt{S}=10.54~\mathrm{GeV}$, corresponding to the $\Upsilon(4S)$ resonance. 

At DESY, measurements in the range covering approximately $10\lesssim\sqrt{S}\lesssim44~\mathrm{GeV}$
were first performed at the DORIS storage ring by PLUTO~\cite{PLUTO:1980mrm} and ARGUS~\cite{ARGUS:1989orf,ARGUS:1992dxx}
and continued at higher energies in the 1980s at the PETRA collider with the TASSO
\cite{TASSO:1980dyh,TASSO:1982bkc, TASSO:1984nda, TASSO:1988jma, TASSO:1989jyt}, JADE~\cite{JADE:1983czq} and CELLO~\cite{CELLO:1989byk}.

At KEK, data were initially collected by TOPAZ \cite{TOPAZ:1994voc} at the TRISTAN 
$e^+e^-$ collider operating at c.m.s.~energies of up to about $\sqrt{S}=60~\mathrm{GeV}$.
Like at SLAC, data taking continued until 2010 with much higher statistics but at a lower energy near
$\sqrt{S}=10.58~\mathrm{GeV}$ at the KEKB $B$-factory by the 
Belle collaboration \cite{Belle:2013lfg,Belle:2015hut}. They sill collect data at the intensity frontier,
but now as BELLE-II at the upgraded SuperKEKB facility.
Due to the extremely high luminosities achieved at both $B$-factories at SLAC and KEK, 
the Belle and BaBar measurements currently provide the most precise constraints on FFs,
only limited by systematic uncertainties.

In addition to SLD, the principal information on SIA at the highest energies so far, 
comes from the four experiments ALEPH \cite{ALEPH:1994cbg}, DELPHI \cite{DELPHI:1998cgx},
L3~\cite{L3:1992pbe,L3:1994gkb}, and OPAL \cite{OPAL:1994zan} at the LEP $e^+e^-$ circular collider, 
which operated at CERN until 2000 predominantly at $\sqrt{S}=M_Z$. Towards the end,
some additional measurements were performed at c.m.s.~energies above the $Z$-boson 
mass \cite{OPAL:2002isf, DELPHI:2000ahn}. 
While most of the LEP data correspond to SIA, a limited set of measurements is also available 
for hadron-in-jet production \cite{L3:1995rnv, ALEPH:1999udi, OPAL:2004prv}, 
see also Sec.~\ref{sec:injet}.

BES-III at IHEP, the only other still operating experiment apart from BELLE-II, has extended 
SIA measurements down to $\sqrt{S} \approx 2.23-3.67$ GeV \cite{BESIII:2022zit, BESIII:2025mbc}.
These data provide a new low-scale lever arm in studying and exploiting the scale evolution
of FFs but they are certainly at the edge of applicability of pQCD methods and factorization.
Hence, their analysis in terms of FFs requires particular care, also due to potentially
large hadron-mass effects and other power-suppressed contributions.

Besides fully inclusive SIA measurements discussed so far, flavor-tagged data, in which the primary 
$q\bar{q}$ pair produced in the reaction is identified under some assumptions
by experimental means, 
have also been reported by the TPC \cite{TPCTwoGamma:1986evp}, SLD \cite{SLD:2003ogn}, 
DELPHI \cite{DELPHI:1998cgx}, and OPAL \cite{OPAL:1994zan} collaborations. 
The idea is roughly to determine the probabilities for having an initial $b\bar{b}$ or $c\bar{c}$ pair
in SIA by utilizing the characteristic decay topologies and lifetimes of heavy quarks in a tagged-jet
and then look for a specific hadron in the away-side jet. In a LO picture for SIA, one can then 
infer the rate of hadrons originating from a bottom or charm quark. The remaining part of the 
cross section is then assumed to be due to the sum of light $u,\,d,\,s$ quark fragmentation.
Despite obvious difficulties and limitations in interpreting these results 
in a higher-order pQCD framework, heavy-flavor tagged data are used in all fits of FFs 
due to the lack of any better constraints, see also Sec.~\ref{subsec:heavyquarks} for
the treatment of heavy flavors in the scale evolution of FFs.
The OPAL collaboration \cite{OPAL:1994zan} went one step further by providing also 
fully flavor-separated $u\bar{u}$, $d\bar{d}$, and $s\bar{s}$ samples, that, however,
heavily rely on the use of MC simulations. Because of the strong model dependence,
there is no unambiguous theoretical interpretation of these data, and, hence, 
they are of very limited use.

No updates to the existing suite of SIA data are expected anytime soon. Any progress depends 
on future $e^+e^-$ collider projects (FCCee, ILC, CLIC, CEPC) coming to fruition.

\subsection{Identified particles in semi-inclusive DIS}
%
In the case of SIDIS measurements, the most relevant data sets have been obtained 
at DESY, CERN, and JLab. As discussed in Sec.~\ref{sec:sidis-theory}, these results
are indispensable for the flavor and quark/antiquark separation of FFs in global analyses.

The HERA accelerator at DESY operated for 15 years until 2009 and
hosted two large-scale collider experiments, H1 and ZEUS.
It collided a proton beam of up to $920\,\mathrm{GeV}$ with either a $27.5\,\mathrm{GeV}$
electron or positron beam, corresponding to a c.m.s.~energy of $\sqrt{S}=318\,\mathrm{GeV}$. 
The lepton beam in HERA became automatically transversely polarized during storage
through the Sokolov–Ternov effect. The direction of polarization could be controlled 
with spin rotators for each experiment.
HERA was also home to the HERMES experiment, which scattered the lepton beam off an
internal gas target such as hydrogen or deuterium at a much lower energy of 
$\sqrt{S}=7.2\,\mathrm{GeV}$ than H1 and ZEUS. Yet, thanks to 
its particle identification (PID) capabilities, HERMES delivered precise
measurements of SIDIS multiplicities (\ref{eq:sidis-mult})
for $\pi^\pm$ and $K^\pm$ \cite{HERMES:2001ghm,HERMES:2012uyd}
which still play a crucial role in the extraction of FFs today, albeit with discussions
about the applicability of pQCD in the covered kinematical range.
The availability of polarized gas targets also allowed for pioneering measurements 
of spin asymmetries in SIDIS and for related processes.
The main focus of the two collider experiments was on inclusive DIS with a 
broad kinematic coverage up to large values of $Q$, but due to their limited PID 
only SIDIS data with unidentified charged hadrons, $K^0_S$, $\Lambda/\overline{\Lambda}$ 
have been published by H1~\cite{H1:2007ghd,H1:2009lef} and 
ZEUS~\cite{ZEUS:1999ies, ZEUS:2009rju, ZEUS:2010mrq, ZEUS:2011cdi}.

CERN has a long legacy of DIS-type fixed-target experiments using naturally polarized
muon beams from the SPS with energies from 100 GeV to 280 GeV.
The EMC experiment measured charged pion and unidentified charged hadron
multiplicities in the 1980s using a liquid deuterium target  
at $\sqrt{S}\approx29\,\mathrm{GeV}$ \cite{EuropeanMuon:1985nbu}
and both liquid hydrogen and deuterium targets in the range 
$\sqrt{S}\approx13-29\,\mathrm{GeV}$ \cite{EuropeanMuon:1991sne}.
CERN's efforts on SIDIS concluded in 2022 with the final run of the
COMPASS experiment that utilized liquid hydrogen and polarizable solid lithium deuteride
and ammonia targets and a 160 GeV muon beam corresponding to $\sqrt{S}\approx17\,\mathrm{GeV}$.
COMPASS was substantially more precise than EMC and 
obtained SIDIS multiplicities for $\pi^\pm$, $K^\pm$, and unidentified 
charged hadrons. Their latest results, which implement also improved radiative QED 
corrections, can found in \cite{COMPASS:2024gje,COMPASS:2025bfn} and are routinely 
analyzed in current fits of FFs.

Data are still being collected at JLab utilizing the continuous electron beam of
CEBAF with energies up to 10.6 GeV available for the experimental halls A, B, and C.
In particular, the spectrometer experiments of Hall C have released precise $\pi^\pm$
SIDIS multiplicities for proton and deuterium targets 
\cite{Bhatt:2024prq,HallCSIDIS:2025plr} that provide a low-energy complement to other SIDIS data sets.
Because of the very moderate $\sqrt{S}\approx4.5\,\mathrm{GeV}$
the incorporation of these data in global analyses of FFs has been reluctant so far as
it would require finite target- and hadron-mass corrections and a careful consideration of
higher-twist effects and all-order resummations of threshold logarithms.

We do not expect to see any significant new results in SIDIS 
pertinent to the determination of collinear FFs before
the advent of EIC operations in the mid-to-late 2030s.

\subsection{Identified particles in hadronic collisions}
%
As discussed in Secs.~\ref{sec:pp-theory} and \ref{sec:injet}, measurements of 
hadron production in $pp$ and $p\bar{p}$ collisions supply the most direct constraints on the 
gluon FF so far, since gluon-initiated subprocesses significantly contribute to the relevant 
cross sections already at LO accuracy. The most relevant data sets for hadroproduction stem from experiments 
being performed at high-energy colliders at Fermilab, BNL, and CERN over the past decades.

At Fermilab, the two large experiments CDF and D0 at the TeVatron,
a $p\bar{p}$ collider operating until 2011 at c.m.s.\ energies of $\sqrt{S}=630$ and
$1800\,\mathrm{GeV}$, both lacked good PID capabilities. High-$p_T$ hadronproduction
data are only available from CDF for unidentified charged hadrons, 
as well as $K_S^0$ and $\Lambda$-baryons \cite{CDF:1988evs,CDF:2005ofj, CDF:2009cxa}. 

At BNL, the versatile RHIC, which ceased operations in 2026, had two large scale 
experiments, PHENIX  \cite{PHENIX:2007kqm,PHENIX:2010hvs,PHENIX:2012koo,PHENIX:2014axc,PHENIX:2014qwb,PHENIX:2015fxo} 
and STAR
\cite{STAR:2006dgg,STAR:2006nmo,STAR:2006xud,STAR:2009vxb,STAR:2012ljf,STAR:2013zyt}, 
that collected a large suite of high-$p_T$ data in $pp$ collisions between $\sqrt{S}=62.4$ 
and 510 GeV mainly at mid-rapidity and for neutral and charged pions but results are 
also available for kaons, $\eta$ mesons, (anti)protons, and $\Lambda/\bar{\Lambda}$ baryons.
In addition, RHIC was the only collider capable of polarizing both of the proton beams
either longitudinally or transversely or running with a large selection of 
different heavy-ion beams. These runs also produced valuable data sets with identified hadrons
that are used for other purposes than fits of FFs.

The available energy range for $pp$ collisions was vastly expanded up to 
$\sqrt{S}=13\,\mathrm{TeV}$ with the start of first collisions at the LHC at CERN in 2009. 
Thanks to their PID capabilities, the ALICE experiment supplies now the most extensive 
data sets for inclusive high-$p_T$ hadron measurements with both identified
charged and neutral hadrons \cite{ALICE:2014juv, ALICE:2017ryd, ALICE:2019hno, ALICE:2020jsh, ALICE:2021est, ALICE:2024vgi}. 
In addition, the two major LHC experiments, ATLAS and CMS, provide high-precision inclusive spectra for
unidentified charged hadrons over wide range of transverse momentum and pseudo-rapidity 
\cite{ATLAS:2010jvh, CMS:2011mry, CMS:2012aa} as well as hadron-in-jet data and hadron production 
in photon- and Z-tagged events 
\cite{ATLAS:2011myc,ATLAS:2018bvp, ATLAS:2019dsv,ATLAS:2019rqw,ATLAS:2020wmg, CMS:2018mqn, CMS:2021otx}. 
Complementary data are provided by the LHCb collaboration which specializes
on measurements at very large rapidity taken with their single-arm forward spectrometer. 
So far they have reported data for both inclusive hadron production 
\cite{LHCb:2021abm,LHCb:2023iyw} as well as hadron-in-jet events including PID
with an associated tagged $Z$ boson \cite{LHCb:2019qoc,LHCb:2022rky}.

We note that older, legacy data such as those obtained in the 1980s at CERN's early SPS collider
experiments UA1~\cite{Bocquet:1995jr, UA1:1989bou} and UA2~\cite{UA2:1984ida}, 
at $\sqrt{S}=200-900\,\mathrm{GeV}$ and $540\,\mathrm{GeV}$, respectively, 
as well as fixed-target results from, for instance, the Fermilab E706 collaboration 
\cite{FermilabE706:2002wtp} at $\sqrt{S}\approx31.5\,\mathrm{GeV}$
are usually no longer considered in current fits of FFs. On the one hand, they largely cover
only relatively small values of $p_T$ and on the other hand they are much less precise than
more modern measurements. In addition, fixed-target results are known to receive large 
higher-order QCD corrections \cite{Arleo:2008zd} and likely even require the inclusion of 
threshold resummations \cite{deFlorian:2005yj,Hinderer:2018nkb} and power corrections. 
A similar remark applies to the data collected at BNL's RHIC at $\sqrt{S}=200\,\mathrm{GeV}$,
very forward rapidity and low $p_T$ by both STAR \cite{STAR:2006dgg} and BRAHMS \cite{BRAHMS:2007tyt}.
Here, one now uses corresponding data from LHCb at much higher values of $\sqrt{S}$ and $p_T$ and,
hence, in a kinematic region where pQCD should be safely applicable.

\section{What do we know about fragmentation functions}\label{chap4:fits}
%
In this final chapter, we reunite the theoretical formalism outlined in Chap.~\ref{chap2:theory}
and the experimental results listed in Chap.~\ref{chap3:data} by
reviewing how to bring a global QCD analysis of FFs into fruition.
First, we discuss how symmetries and sum rules may help to reduce the space of
possible functions describing the set of parton-to-hadron FFs in a fit. Next, we present
different strategies used in current global analyses to parametrize the FFs at some initial 
scale $\mu_0$, to determine them from a fit to data, and to assess the quality of the fit.
We also cover the different methodologies to estimate the uncertainties of the extracted FFs.
Finally, we compare the individual quark flavors, the singlet combination, and the gluon 
from four (three) state-of-the-art fits of parton-to-pion FFs at (N)NLO accuracy,
including their respective uncertainty estimates.

\subsection{Constraints from sum rules, isospin and charge symmetry}
\label{subsec:symmetries}
To limit the number of independent FFs in a QCD analysis, one can resort to
charge conjugation and/or isospin symmetry. While $C$-parity is an exact
symmetry of QCD and links the fragmentation of a parton $i$ into a hadron $h$
to that of the corresponding anti-parton $\bar{i}$ into the anti-hadron $\bar{h}$,
i.e., $D^h_i(z,\mu) = D^{\bar{h}}_{\bar{i}}(z,\mu)$,
isospin symmetry is only approximate and broken by 
the $u-d$ mass difference and electroweak interactions.
Assuming that up and down (anti)quarks form an SU(2) doublet, 
one can relate their FFs into hadrons that are isospin partners
of each other, e.g.\ $D_u^{\pi^{+}}=D_d^{\pi^{-}}$ or 
$D_u^{K^{+}}=D_d^{K^{0}}$. 
Combining these constraints one arrives, in case of pions, at
only one set of so-called favored and one set of unfavored FFs,
$D_u^{\pi^{+}}=D_d^{\pi^{-}}=D_{\bar{d}}^{\pi^{+}}=D_{\bar{u}}^{\pi^{-}}$
and $D_u^{\pi^{-}}=D_d^{\pi^{+}}=D_{\bar{d}}^{\pi^{-}}=D_{\bar{u}}^{\pi^{+}}$, 
respectively.
While SU(2) isospin symmetry is expected to work at the $\mathcal{O}(1\%)$ level, 
an extended SU(3) flavor symmetry, which includes also
strange (anti)quarks, likely holds only to about $\mathcal{O}(20-30\%)$
due to the much larger strange quark mass. 
Such magnitude of flavor symmetry violations is also found from fits to data
that do not impose exact isospin symmetry, see, e.g., \cite{deFlorian:2007aj}.
Recent JLab data for $\pi^{\pm}$ SIDIS multiplicities \cite{Bhatt:2024prq}
suggest a larger than expected breaking of SU(2) isospin symmetry for 
unfavored FFs with decreasing energy of the collision.
It remains to be seen if these low-energy data can be accommodated in
future global analyses as factorization might not apply.
We note that potential isospin breaking due to weak decays 
has been estimated in \cite{Chen:2021hdn}.

\subsection{Strategies to extract fragmentation functions and their uncertainties
from data}\label{chap4:subsubsec2}
%
While the scale evolution of FFs is governed by pQCD, see Sec.~\ref{sec:scaledep}, the distributions themselves 
still need to be specified at some arbitrary value of the factorization scale $\mu_0$. 
The tasks by which these initial conditions are determined translate into first parameterizing 
the set of individual parton-to-hadron FFs at $\mu_0$ and then choosing the details of the theoretical framework
such as the perturbative order, the specifics of the solution of the evolution equations including the
treatment of heavy flavors and, perhaps, the inclusion of all-order resummations. Next, one needs to select
the data sets for a given hadron species $h$ which are going to be used in the analysis and impose certain
kinematic cuts on them to limit unwanted contributions from, say, power corrections. 
The set of parameters describing the FFs $D_i^h(z,\mu_0)$ that optimizes the description of the 
experimental data from many different sources is then determined by means of a global analysis. 
In this step, one also needs to propagate experimental errors and, perhaps, sources of theoretical ambiguities 
to the resulting set of FFs to determine not only a best fit but also to arrive at a faithful uncertainty estimate.
To this end, a viable criterion that measures the quality of a fit in a high-dimensional parameter space
and in the presence of many nonuniform sources of uncertainties.
Needless to say, that fits of FFs performed by different groups often deviate in some or even many of the steps 
just outlined, which complicates a comparison of the results.

An often adopted, simple but flexible 
initial functional-form for the hadronization of a parton $i$ 
into a hadron $h$ at scale $\mu_0$
is given by \cite{deFlorian:2014xna,Moffat:2021dji,Gao:2024dbv}
\begin{equation} 
\label{eq:ff_param}
D_i^h(z,\mu_0) = N_i\,z^{\alpha_i}(1-z)^{\beta_i} f_i(z,\theta_i)\,,
\end{equation}
which forces the distribution to vanish in the limit $z\rightarrow 1$ as it should.
Its behavior at small $z$ is determined by the value of $\alpha_i$.
The term $f_i(z,\theta_i)$ corresponds to a function of $z$ and additional parameters 
collectively denoted as $\theta_i$. 
For a simple enough choice of $f_i$, e.g.~polynomial, the ansatz (\ref{eq:ff_param}) can be
analytically expressed in Mellin space by Euler beta functions.
As explained in the previous section, the sizable number of free parameters $N_i$,
$\alpha_i$, $\beta_i$, and $\theta_i$ in (\ref{eq:ff_param}) to be determined by a global fit
for each flavor $i$ can be significantly reduced by charge and isospin symmetries
if applicable.
The level of complexity of the functional form adopted in (\ref{eq:ff_param})
depends on the availability and quality of the data sets used to constrain them.
The less precise they are, the more ad-hoc assumptions on the parameter space
need to be made to arrive at a meaningful fit.
These choices necessarily constitute some hard to quantify bias in both the obtained FFs 
and their uncertainties.
To circumvent these issues introduced through the choice of a specific fixed 
functional form (\ref{eq:ff_param}), the use of neural networks (NN) to parametrize the 
initial set of FFs has been proposed and successfully explored in recent years
\cite{Bertone:2017tyb,Khalek:2021gxf,AbdulKhalek:2022laj,Soleymaninia:2026xjq,Soleymaninia:2026edd}. 
Here, a training set of data is used to adjust the many internal weights of the network
and a validation step checks how well the model already can cope with unseen data. 
It is crucial, however, to avoid over-learning the NN, otherwise the 
extracted FFs will include all random features of the fitted data sets.

While the complementarity of measurements of SIA, SIDIS, and in $pp$ collisions 
in the extraction of FFs has been highlighted already to some degree in Sec.~\ref{sec:sia-theory}, \ref{sec:sidis-theory}, 
and \ref{sec:pp-theory}, it is worth summarizing some key features regarding the constraints such data impose in global analyses.
The most important constraints to determine FFs come from electron-positron annihilation into charged hadrons, 
which provide the cleanest access to FFs since no other non-perturbative quantities such as PDFs 
are involved in the theoretical description of the cross section. 
Historically, SIA measurements motivated early attempts to model FFs \cite{Field:1977fa,Field:1989uq}
and were still the only source of data included in first determinations of FFs
at NLO accuracy \cite{Binnewies:1994ju, Kniehl:2000fe, Albino:2005gd, Kretzer:2000yf}. 
Fits analyzing exclusively $e^+e^-$ data have, however, considerable shortcomings as they need to impose
several ad-hoc assumptions. Apart from not being able to differentiate 
between quark and anti-quark fragmentation,
it happens by coincidence that the electroweak couplings $\hat{e}_q$ of up-type and down-type quarks 
to the $Z$ boson in Eq.~(\ref{eq:FTFL})
become almost equal at $Q\approx M_Z$ where the bulk of the CERN and SLAC data resides.
While this translates into the still best constraint on the flavor singlet combination
(\ref{eq:singlet}) of FFs, nothing can be inferred on its decomposition into individual flavors. 
To some degree, this is nowadays mitigated by the use of precise SIA data taken 
at B-factories at lower c.m.s.~energies, in which the contributions from different quarks are weighted 
according to their electric charge, thus probing a different combination of FFs. 
In principle, the flavor separation can be further improved by the inclusion of flavor-tagged 
SIA data which, however, lack an unambiguous theoretical interpretation, see also Chap.~\ref{chap3:data}.
In addition, constraints on the gluon fragmentation coming solely from $e^+e^-$ data 
are rather weak as contributions to SIA with a fragmenting gluon first appear in pQCD
at the NLO accuracy and the lever-arm in scale is too limited to utilize scaling-violations.
More differential observables in $e^+e^-$ collisions can provide a stronger handle on the gluon-to-hadron FF.
In particular, the production of identified hadrons inside jets in $e^+e^-$ annihilation allows one 
to directly probe $D_g^h$ if events with at least 3 jets are considered. 
As opposed to inclusive hadron production, these events do require the emission of an 
additional gluon already at LO accuracy. 
Data for hadron-in-jet production in $e^+e^-$ has been included in fits of FFs 
only in \cite{Gao:2024dbv,Gao:2025bko}.

%
\begin{figure}[ht!]
\centering
\includegraphics[width=.9\textwidth]{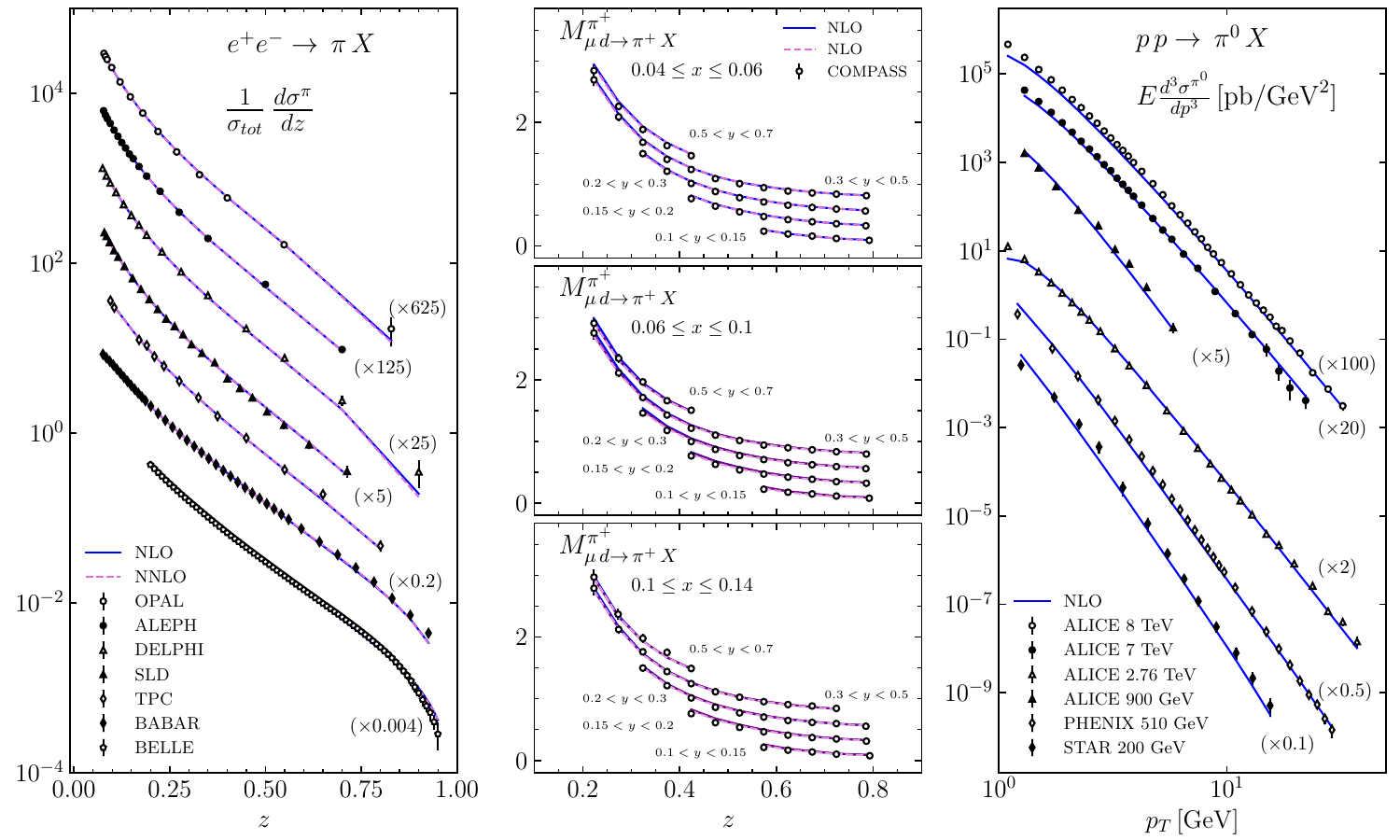}
\caption{Representative comparison between selected data sets used in the determination of parton-to-pion FFs 
and the corresponding theoretical predictions obtained with the extracted FFs. 
Results for pion production in SIA (left panel), SIDIS (central panels), and $pp$ collisions (right panel) 
are presented. In the case of SIA and SIDIS, the theoretical predictions are obtained both at 
NLO (solid lines) and NNLO (dashed lines) accuracy using the set of FFs from \cite{Borsa:2022vvp}. 
For $pp$, the cross sections for neutral pion production are presented up to NLO accuracy 
using the FFs from \cite{Borsa:2021ran}.}
\label{fig:xsecs}
\end{figure}
The limitations just mentioned can be overcome by considering one-particle inclusive measurements 
coming from deep-inelastic lepton-nucleon scattering and proton-proton collisions in global analyses.
These measurements not only probe fragmentation in a complementary energy regime, 
see Fig.~\ref{fig:FFs_coverage}, but also account for the varying contributions of individual 
parton flavors in the hadronization process.
Specifically, for SIDIS data, weighting the contributions of different initial-state partons 
by their respective PDFs significantly improves both quark-antiquark and flavor separation.
The first extraction of FFs based on the combined use of SIA and SIDIS data 
was performed in \cite{Kretzer:2001pz} and is by now standard in all recent global analyses.
Thanks to dominant contributions from gluon-induced fragmentation already at
LO accuracy, data obtained in hadronic collisions are by far the best source 
of information on $D_g^h$ despite the much more involved theoretical framework.
Fully global QCD analyses including all three sources of data at NLO accuracy
were first performed in \cite{deFlorian:2007aj,deFlorian:2007ekg}. 

The typical quality of current parton-to-pion FFs in describing selected data sets used 
for their extraction is illustrated in Fig.~\ref{fig:xsecs}.
Results for pion production in SIA (left panel) and SIDIS (central panels) 
are compared to representative theoretical predictions obtained both at 
NLO (solid lines) and NNLO (dashed lines) accuracy using the set of FFs from \cite{Borsa:2022vvp}.
Although hardly visible, NNLO corrections lead to some 
improvements in fits to SIDIS data, mainly at the smallest measured values of $z$.
Because the NNLO corrections for hadron production in $pp$ collisions \cite{Czakon:2025yti} are 
not yet available in a format easily usable for global analyses of FFs, 
the right panel of Fig.~\ref{fig:xsecs} shows only NLO results based on 
the fit of FFs in \cite{Borsa:2021ran}.
For more detailed comparisons between particular data sets and theory, including also
estimates of uncertainties, e.g., from scale or PDF variations, 
we have to refer the reader to the publications regarding the different available set 
of FFs listed in Tab.~\ref{tab:fits} below.

The parameters describing the FFs at the initial scale $\mu_0$ in Eq.~(\ref{eq:ff_param}) are 
determined by minimizing some figure of merit, traditionally the correlated $\chi^2$
between $N$ data points and theory, 
\begin{equation}\label{eq:chi2}
    \chi^2= \sum_{i,j}^{N}\left(T_i-E_i\right)\mathrm{cov}^{-1}_{ij}\left(T_j-E_j\right)\,,
\end{equation}
with $E_i$ representing the central value of a given observable, 
$T_i$ the corresponding theoretical estimate for a given set of parameters 
in Eq.~(\ref{eq:ff_param}), and $\mathrm{cov}_{ij}^{-1}$ denoting the 
inverse covariance matrix, 
\begin{equation}\label{eq:cov_matrix}
    \mathrm{cov}_{ij}\equiv \sigma_i^{\mathrm{unc}} \sigma_j^{\mathrm{unc}}\delta_{ij}
    +\sum_k \sigma_{k,i}^{\mathrm{corr}} \sigma_{k,j}^{\mathrm{corr}}\,,
\end{equation}
where $\sigma_i^{\mathrm{unc}}$ and $\sigma_{k,i}^{\mathrm{corr}}$ correspond to the 
various sources of uncorrelated and correlated uncertainties, respectively. 
Experimental normalization uncertainties can either be included as a fully 
correlated source of error in each of the data sets, such as 
in \cite{Khalek:2021gxf, Gao:2024dbv}, or used to 
determine analytical normalization coefficients, see \cite{deFlorian:2014xna}.  
Similarly, theoretical uncertainties due to missing higher orders or PDFs
can be also included as additional terms in the covariance matrix (\ref{eq:cov_matrix}).

Each evaluation of $\chi^2$ in Eq.~(\ref{eq:chi2}) requires evolving the FFs for a given 
set of parameters to the scale $\mu$ of each of the $N$ data points and obtaining the corresponding
theoretical predictions. Consequently, the minimization procedure relies critically on 
efficiently solving the evolution equations and calculating the relevant cross sections. 
As discussed in Sec.~\ref{subsec:solving-rge}, this can be achieved both in 
Mellin moment and in $z$ space by pre-computing all numerically expensive but 
parameter-independent parts of the calculation and storing them in large grids.
The minimization of $\chi^2$ itself is usually achieved by applying a gradient descent method
as implemented, for instance, in the \texttt{Minuit} algorithm \cite{James:1975dr}
or variants thereof.
This represents a considerable additional computational burden that increases with the size
of the parameter space. Therefore, other non-gradient or genetic algorithms have
been used in global analyses based on NN.

Besides determining the optimal set of parameters 
$\{N_i,\alpha_i,\beta_i,\theta_i\}$ for all partons $i$ in Eq.~(\ref{eq:ff_param}), 
propagating uncertainties to the extracted FFs is an integral part of a global analyses. 
Three main approaches have been established to estimate fit uncertainties: the Lagrange 
multiplier method, the Hessian method and its iterative implementations, 
and Monte Carlo replica sampling. 
We will briefly examine the core of each method in the remainder of this section.

The Lagrange multiplier (LM) method \cite{Pumplin:2000vx, Stump:2001gu} provides a direct 
and robust way to probe the uncertainty of any observables or quantities of interest
$O_i$ depending on FFs, including the FFs themselves. 
It relates variations of the $O_i$ to changes in the figure of merit 
(\ref{eq:chi2}) used to judge the goodness of the fit to data.
Schematically, this amount to minimizing the function 
\begin{equation}\label{eq:LM}
    \Phi(\lambda_i,\{\Theta_j\}) = \chi^2(\{\Theta_j\})+ \sum_i \lambda_i \, O_i(\{\Theta_i\})
\end{equation}
with respect to the set of parameters $\{N_i,\alpha_i,\beta_i,\theta_i\}$ 
describing the FFs, here collectively denoted as $\{\Theta_i\}$,
for fixed values of the Lagrange multipliers $\lambda_i$.
Minimizing $\Phi(\lambda_i , \{\Theta_j\})$ generally shifts the fitted observables $O_i$ 
away from their best-fit values (obtained for $\lambda_i=0$) 
at the expense of an increase in $\chi^2$.
From a series of global fits for different values of $\lambda_i$, the 
$\chi^2(\{\Theta_j\})$-profile depending on different values of $O_i$ can be 
mapped out precisely. The uncertainty of the observables $O_i$ can then be estimated 
as its range of variation compatible with a prescribed, empirical 
increase or tolerance $T=\sqrt{\Delta\chi^2}$ above the minimum. 
While the LM method avoids any approximations or assumptions about the dependence 
of the $\chi^2$-hypersurface on the parameter space $\{\Theta_i\}$
or on how to propagate uncertainties to a given observable, it requires an extensive 
extra amount of $\chi^2$-minimizations. 
In addition, the resulting scan is specific to the set of observables considered
and requires access to the full fitting machinery. Hence, the LM method is unsuitable
for end users of FFs.

A computationally more flexible and end user friendly alternative to the LM method is 
provided by the iterative Hessian (IH) approach \cite{Pumplin:2000vx, Pumplin:2001ct}. 
The central idea - and, unfortunately, the main shortcoming - of the method
is to assume a textbook-like quadratic behavior of the $\chi^2$-hypersurface 
around its minimum value.
Any variation of the parameters $\{\Theta_i\}$ then results in a change in
$\chi^2$ given by
\begin{equation}\label{eq:hessian_chi2}
\Delta\chi^2=\sum_{ij}y_i\,H_{ij}\,y_j\,,    
\end{equation}
where $y_i\equiv\Theta_i-\Theta_i^0$ represents the displacement of the $i$-th parameter 
$\Theta_i$ from its best-fit value $\Theta_i^0$ and
$H_{ij}$ corresponds to the Hessian matrix
\begin{equation}\label{eq:hessian_matrix}
H_{ij}=\frac{1}{2}\left.\left(\frac{\partial^2\chi^2}
         {\partial y_i\partial y_j}\right)\right|_{y_k=0}\,.
\end{equation}
Iteratively diagonalizing $H_{ij}$ yields an orthonormal basis of directions 
in parameter space (eigenvectors) within the quadratic approximation that 
can be used to express the $y_i$ in Eq.~(\ref{eq:hessian_chi2}) in terms 
of new uncorrelated parameters $z_i$ which, in addition, are rescaled such 
that $\Delta\chi^2=\sum_i z_i^2$ is the squared distance from the minimum.
This eigenvector representation of the Hessian matrix is extremely 
convenient for computing the uncertainties of the FFs 
and for propagating them to arbitrary observables $O$.
To this end, one constructs two eigenvector sets $S_i^\pm$ of FFs for each
parameter $z_i$ that correspond to positive and negative displacements $z_i=\pm T$
and $z_{j\neq i}=0$.
With these Hessian sets at hand, it is now straightforward to compute the  
symmetric uncertainty $\Delta O$ of any observable $O$ that depends on FFs as 
\begin{equation}\label{eq:hessian_err}
    \Delta O=\frac{1}{2}\left[\sum_i \left[O(S_i^+)-O(S_i^-)\right]^2\right]^{1/2}\,.
\end{equation}
Uncertainty estimates from both the LM and the IH method 
and their interpretation depend on the choice of the tolerance parameter $T$. 
In standard Gaussian statistics, a $1\sigma$ or 68$\%$ confidence level (CL) 
uncertainty corresponds to an increase of $\Delta\chi^2=1$ from the minimum. 
However, the global PDF and FF fitting community routinely chooses a
dynamical tolerance $T\gg 1$ that allows for much larger excursions from the best fit,
motivated by bias in the analysis framework, tensions between experimental data
sets that introduce non-Gaussian uncertainties, and highly non-linear
dependencies within the parameter space. This approach not only complicates 
the comparison of fits and uncertainties across different groups but also 
introduces some level of arbitrariness to the fit results.
A systematic comparison of uncertainties for pion and kaon FFs 
derived from the LM and the Hessian method has been presented 
in \cite{Epele:2012vg}. It is shown that the uncertainties roughly coincide 
if the tolerance for the IH method is set to more than twice 
that of the LM method. This clearly signals a departure from 
the assumed quadratic dependence of the $\chi^2$-profile in the IH approach
and, hence, its limitations.

The Monte Carlo replica method entirely bypasses the need for a tolerance 
criterion and any assumptions about the dependence of the 
$\chi^2$-profile on the fit parameters.
Instead, it represents the uncertainties of the extracted FFs through 
a sufficiently large ensemble of independently fitted replicas. 
For this purpose, $N_{\mathrm{rep}}$ pseudo-datasets are generated by randomizing 
the central values of each measurement based on its quoted uncertainties.
Each pseudo-data set is then fitted independently, yielding a FF set $S^{(k)}$.
The entire ensemble of FF replicas therefore propagates the sampled experimental 
uncertainties through the fitting procedure without 
imposing any assumption on $\chi^2$.
Yet its reliability depends on the flexibility of the parametrization
(fixed functional form (\ref{eq:ff_param}) or NN), 
the quality of the $\chi^2$-minimization (each replica must correspond to 
the global and not a local minimum of $\chi^2$), 
and a sufficiently good replica statistics (a significant computational burden). 
For any observable $O$ depending on FFs, its mean value and standard deviation
can be computed statistically from the ensemble of replicas as
\begin{equation}\label{eq:mc_observable}
    \langle O\rangle=\frac{1}{N_{\mathrm{rep}}}
    \sum_{k=1}^{N_{\mathrm{rep}}} O(S^{(k)})
    \qquad\text{and}\qquad
    \Delta O=\left[\frac{1}{N_{\mathrm{rep}}-1}\sum_{k=1}
    ^{N_{\mathrm{rep}}}\left(O(S^{(k)})-\langle O\rangle\right)^2\right]^{1/2}\,,
\end{equation}
respectively, where $O(S^{(k)})$ corresponds to the observable evaluated 
with the $k$-th replica of FFs.
Besides the conceptually clean propagation of uncertainties, another
asset of the method is the possibility to include additional data without 
performing a new global analysis. This is achieved through Bayesian reweighting
\cite{Ball:2010gb}, where the weight of each replica - originally distributed uniformly - is modified according to how well it reproduces the new data.
The mean value in Eq.~(\ref{eq:mc_observable}) is then replaced 
by a weighted average. Of course, this is statistically meaningful
only if a large number of the initial $N_{\mathrm{rep}}$ replicas 
survive the reweighting procedure.

\subsection{Current global fits}\label{chap1:subsubsec1}
%
The phenomenological determination of collinear parton-to hadron FFs and their inherent uncertainties
has evolved considerably over the last three decades. This progress results from the increased availability 
and precision of measurements sensitive to FFs, see Tab.~\ref{tab:collinear-ff-data}, 
as well as significant advances in the theoretical description of these observables and the methodology used 
in global analyses.
In this section, we present a brief historical overview of various determinations of FFs
and highlight recent developments. A summary of the available sets of FFs is presented in Tab.~\ref{tab:fits}, 
whereas earlier reviews can be found in \cite{Albino:2008gy, Metz:2016swz}.
We mainly focus on results for identified pions, which are most abundantly produced in experiment, 
but also list sets of FFs for heavier mesons and baryons up $\Lambda$ hyperons.
%
\begin{table}[ht!]
\centering
\TBL{\caption{Overview of global fits of parton-to-hadron FFs by different groups. 
The columns indicate the type of data used in the analysis, the perturbative order, 
the employed parametrization (fixed functional form or NN), 
the treatment of uncertainties (LM, IH, or MC), 
the types of hadrons considered, and the corresponding references.}\label{tab:fits}}
{
\scriptsize
\setlength{\tabcolsep}{3pt}
\renewcommand{\arraystretch}{1.13}
\begin{tabular}{L{1.9cm} C{0.55cm} C{0.65cm} C{0.55cm} C{1.5cm} C{1.15cm} C{1.05cm} L{3.3cm} L{4.0cm}}
\toprule
\multirow{2}{*}{Fit} & \multicolumn{3}{c}{Datasets} & \multirow{2}{*}{Order} & \multirow{2}{*}{Param.} & \multirow{2}{*}{Uncert.} & \multirow{2}{*}{Final states} & \multirow{2}{*}{Citation(s)} \\
\cmidrule(lr){2-4}
 & $e^+e^-$ & SIDIS & $pp$ & & & & & \\
\midrule
\multicolumn{9}{l}{\textit{LO/NLO fits}} \\
\midrule
EMC & \no & \yes & \no & LO & \fixed & \no & $\pi^\pm$ & \cite{EuropeanMuon:1985nbu,EuropeanMuon:1989srv} \\
CGMG & \yes & \no & \yes & NLO & \fixed & \no & $\pi^0$ & \cite{Chiappetta:1992uh} \\
BKK & \yes & \no & \no & NLO & \fixed & \no & $h^\pm,\pi^\pm,K^\pm,K_S^0$ & \cite{Binnewies:1994ju,Binnewies:1995pt,Binnewies:1995kg} \\
DSV & \yes & \no & \no & NLO & \fixed & \no & $\Lambda$ & \cite{deFlorian:1997zj} \\
Kretzer & \yes & \yes & \no & NLO & \fixed & \no & $h^\pm,\pi^\pm,K^\pm$ & \cite{Kretzer:2000yf, Kretzer:2001pz} \\
BFGW & \yes & \no & \no & NLO & \fixed & IH & $h^\pm$ & \cite{Bourhis:2000gs} \\
KKP & \yes & \no & \no & NLO & \fixed & \no & $\pi^\pm,K^\pm,p/\bar p$ & \cite{Kniehl:2000fe} \\
AKK08 & \yes & \no & \yes & NLO & \fixed & \no & $\pi^\pm,K^\pm,p/\bar p,K_S^0,\Lambda$ & \cite{Albino:2008fy} \\
HKNS & \yes & \no & \no & NLO & \fixed & IH & $\pi^\pm,K^\pm,p/\bar p$ & \cite{Hirai:2007cx} \\
DSS & \yes & \yes & \yes & NLO & \fixed & LM, IH,\mc & $h^\pm,\pi^\pm,K^\pm,p/\bar p,\eta$ & \cite{deFlorian:2007aj,deFlorian:2007ekg,Aidala:2010bn,deFlorian:2014xna, deFlorian:2017lwf, Borsa:2021ran, Borsa:2023zxk,Aidala:2025kep} \\
LSS & \no & \yes & \no & NLO & \fixed & \no & $\pi^\pm$ & \cite{Leader:2015hna} \\
JAM & \yes & \yes & \no & NLO & \fixed & \mc & $\pi^\pm,K^\pm,h^\pm$ & \cite{Ethier:2017zbq, Sato:2019yez,Moffat:2021dji, Cocuzza:2025qvf} \\
NPC & \yes & \yes & \yes & NLO & \fixed & IH & $\pi^\pm,K^\pm,p/\bar p,h^\pm, \eta, K^0_S, \Lambda$ & \cite{Gao:2024nkz, Gao:2024dbv, Gao:2025bko} \\
\midrule
\multicolumn{9}{l}{\textit{Fits with NNLO variants}} \\
\midrule
ASR & \yes & \no & \no & NNLO & \fixed & \no & $\pi^\pm$ & \cite{Anderle:2015lqa, Anderle:2016czy} \\
NNFF & \yes & \no & \yes$^*$ & NNLO & \nn & \mc & $\pi^\pm,K^\pm,p/\bar p,h^\pm$ & \cite{Bertone:2017tyb,Bertone:2018ecm} \\
SGK & \yes & \no & \no & NNLO & \fixed & IH & $h^\pm$ & \cite{Soleymaninia:2018uiv} \\
SAK & \yes & \no & \no & NNLO & \fixed & IH & $K^0_S,\Lambda$ & \cite{Soleymaninia:2020ahn} \\
SGKS & \yes & \no & \no & NNLO & \fixed & IH & $\pi^\pm, K^\pm, p/\bar p$ & \cite{Soleymaninia:2020bsq} \\
IPM-xFitter & \yes & \no & \no & NNLO & \fixed & IH & $\pi^\pm$ & \cite{Abdolmaleki:2021yjf} \\
BDSSV & \yes & \yes & \no & NNLO & \fixed & \mc & $\pi^\pm$ & \cite{Borsa:2022vvp} \\
MAPFF & \yes & \yes & \no & NNLO & \nn & \mc & $\pi^\pm,K^\pm$,$\Lambda$ & \cite{Khalek:2021gxf,AbdulKhalek:2022laj,Bertone:2026fnm} \\
FF24-$K_S^0$ & \yes & \no & \no & NNLO & \nn & \mc & $K_S^0$ & \cite{Soleymaninia:2024jam} \\
NPC & \yes & \yes & \no & NNLO & \fixed & IH & $\pi^\pm,K^\pm$ & \cite{Gao:2025hlm} \\
HAPS & \yes & \yes & \no & NNLO & \nn & \mc & $h^\pm$ & \cite{Soleymaninia:2026xjq,Soleymaninia:2026edd} \\
\bottomrule
\end{tabular}
}{}
\end{table}

As previously mentioned, most of the early NLO analyses of FFs, including the BKK
\cite{Binnewies:1994ju}, Kretzer \cite{Kretzer:2000yf}, KKP \cite{Kniehl:2000fe}, 
and BFGW \cite{Bourhis:2000gs} sets, relied exclusively on SIA data, resulting in
a lack of flavor separation and largely unconstrained gluon fragmentation.
Early attempts to overcome these limitations by supplementing the fit with complementary
observables were made in \cite{Chiappetta:1992uh, Bourhis:2000gs} and \cite{Kretzer:2001pz}, 
which considered data on hadroproduction and SIDIS, respectively.
Beyond expanding the analyzed data sets past SIA, methodological improvements 
focused on the systematic treatment of FF uncertainties, first established 
through the IH method in \cite{Bourhis:2000gs, Hirai:2007cx}.

The DSS group performed the first truly global extraction of FFs, simultaneously fitting 
measurements for SIA, SIDIS, and inclusive high-$p_T$ hadron production in $pp$ collisions 
\cite{deFlorian:2007aj, deFlorian:2007ekg}. They also exploited for the first time
the LM technique to estimate uncertainties of FFs.
These analyses were subsequently updated to include the $e^+e^-$ data from B-factories
as well as $pp$ data from the LHC \cite{deFlorian:2014xna, deFlorian:2017lwf,
Borsa:2021ran,Borsa:2023zxk}. Shortly after the original DSS fit, the AKK08 analysis 
\cite{Albino:2008fy} combined SIA with RHIC and TeVatron data. In addition, they investigated 
the impact of hadron mass corrections as well as large-$z$ threshold resummations on the
quality of the fit.
Other noticeable developments at that time include the first NNLO extraction of FFs 
\cite{Anderle:2015lqa}, which was solely based on pion data in SIA due 
to the lack of NNLO corrections for other relevant processes, 
supplemented by a study of all-order resummations at small-$z$ \cite{Anderle:2016czy}.

Since the last comprehensive review of FFs in 2016 \cite{Metz:2016swz}, several groups 
introduced more flexible and novel methodologies for the determination of FFs 
and, in particular, their uncertainties.
NNFF1.0 \cite{Bertone:2017tyb} determined sets of FFs for $\pi^\pm$, $K^\pm$, and $p/\bar p$ 
up to NNLO accuracy from SIA data, using - for the first time - NN parametrizations rather
than fixed functional forms and MC uncertainties.
Its extension NNFF1.1h \cite{Bertone:2018ecm} used a Bayesian reweighting procedure for 
single-inclusive charged hadron spectra from the TeVatron and LHC to extract a set of 
FFs for unidentified hadrons and to improve the constraints on the gluon FF. 
NN based parametrizations were also adopted in the MAPFF1.0 analysis \cite{Khalek:2021gxf}, 
which combined SIA with SIDIS data to determine charged pion FFs at NLO accuracy.
Yet another strategy was proposed and pursued by the JAM collaboration: 
to simultaneously determine PDFs and FFs. In the JAM20-SIDIS analysis \cite{Moffat:2021dji}, 
the unpolarized PDFs are constrained by inclusive DIS and Drell–Yan data, 
while the $\pi^\pm$, $K^\pm$  and charged hadron FFs are fitted at NLO accuracy 
using SIA and SIDIS data.
Such a unified but more challenging procedure certainly helps to reduce the bias 
introduced by using a specific set of PDFs in an analysis of FFs. 
In addition, SIDIS data should also improve the determination of PDFs, 
particularly for weakly-constrained sea-quark distributions 
\cite{Borsa:2017vwy, Sato:2019yez}, as they are not included in standard PDF fits.
The framework is still incomplete though, as it lacks the inclusion of any hadron 
collider data, which are technically and computationally the most demanding once to
analyze but provide, on the other hand, vital constrains on both the gluon PDF and FF.
The landscape of fits at NLO accuracy was completed by the NPC23 analysis, which currently 
provides one of the most comprehensive determinations of identified charged- and neutral-hadron 
FFs \cite{Gao:2024dbv,Gao:2025bko}, jointly fitting SIA, SIDIS, and inclusive-hadron 
production data from RHIC and the LHC, including measurements of photon-tagged, $Z$-tagged, 
and hadron-in-jet fragmentation results. Theoretical uncertainties associated with the scale 
dependence at NLO accuracy are also included in their fit as an additional contribution 
to the covariance matrix.

The past few years have also witnessed major breakthroughs in the foundational 
theoretical framework for FFs with the computation of the full NNLO QCD corrections 
for SIDIS and high-$p_T$ hadron production, as was already mentioned in Sec.~\ref{sec:sidis-theory} 
and Sec.~\ref{sec:pp-theory}, respectively. This finally allows all groups to push their
analysis framework consistently to NNLO accuracy in the future.
In the meantime, approximate NNLO corrections for SIDIS derived from known threshold resummations
were utilized in \cite{Borsa:2022vvp} to perform the first global extraction of pion FFs 
at NNLO accuracy based on SIA and SIDIS data. 
A similar analysis for both pion and kaon FFs within in NN framework was presented
by the MAP collaboration \cite{AbdulKhalek:2022laj}.
Interestingly enough, both analyses reported a deterioration in the description of SIDIS data
at NNLO accuracy as long as low-$Q^2$ data were included in the fit. 
The exact NNLO expressions for SIDIS were first utilized by NPC in a determination of 
pion kaon FFs from SIA and SIDIS data \cite{Gao:2025hlm}. 
An equally good description of SIDIS measurements at both NLO and NNLO accuracy was obtained.
A similar NNLO analysis was recently performed by the HAPS collaboration 
\cite{Soleymaninia:2026edd, Soleymaninia:2026xjq}.

\begin{figure}[h]
\centering
\includegraphics[width=.9\textwidth]{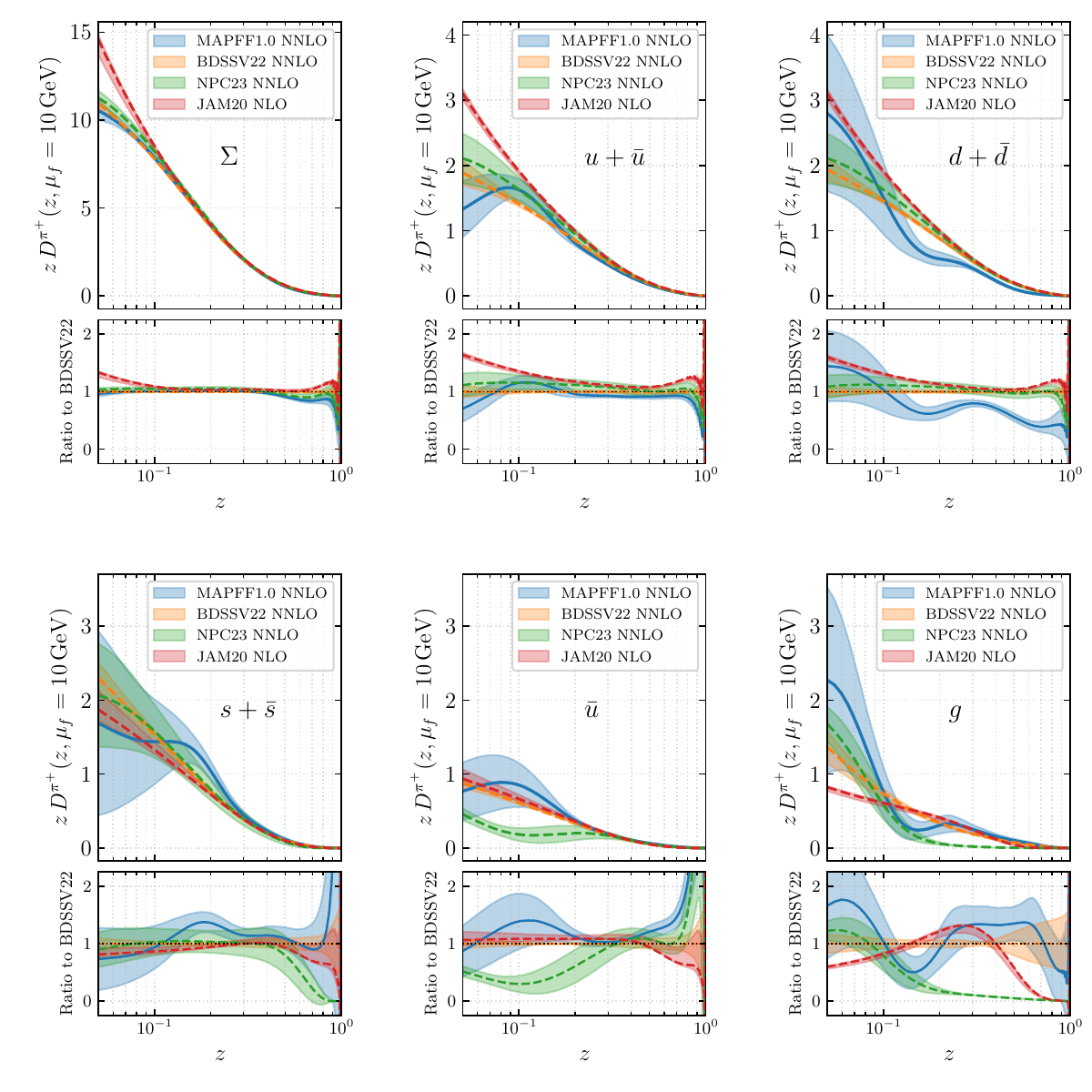}  
\caption{Comparison of the parton-to-pion FF corresponding to the singlet combination $\Sigma$, 
$u+\overline{u}$, $d+\overline{d}$, $s+\overline{s}$, $\overline{u}$, and the gluon 
at $\mu_f=10\,\mathrm{GeV}$. The different curves correspond to the results 
of the analyses by MAPFF1.0~\cite{AbdulKhalek:2022laj}, BDSSV22 \cite{Borsa:2022vvp}, 
and NPC23 \cite{Gao:2025bko} at NNLO accuracy, as well as JAM20-SIDIS \cite{Moffat:2021dji} 
at NLO accuracy. The lower insets depict the ratio relative to the FF combination from BDSSV22.
The shaded bands correspond to $1\sigma$ uncertainty estimates for each fit.}
\label{fig:FFs_mixed}
\end{figure}

A comparison between different sets of parton-to-pion FFs obtained in recent global fits 
is presented in Fig.~\ref{fig:FFs_mixed}. In particular, we show results, including
estimates of uncertainties (shaded bands), for the combinations 
$D_{\Sigma}^{\pi^+}$, $D_{u+\bar u}^{\pi^+}$, 
$D_{d+\bar d}^{\pi^+}$, $D_{s+\bar s}^{\pi^+}$, the unfavored FF $D_{\bar u}^{\pi^+}$, 
and the gluon FF $D_{g}^{\pi^+}$ at $\mu_f=10\,\mathrm{GeV}$.
All fits are based exclusively on SIA and SIDIS data. The MAPFF1.0 \cite{AbdulKhalek:2022laj}, 
BDSSV22 \cite{Borsa:2022vvp}, and NPC23 \cite{Gao:2025bko} analyses are performed 
at NNLO accuracy, whereas the JAM20-SIDIS \cite{Moffat:2021dji} results are obtained at NLO.
To facilitate the comparison between the different fit results, the lower insets in each panel 
display the ratio relative to the FF combination from BDSSV22.
%
%

Although making direct comparisons between different fits is very subtle and requires careful 
consideration of differences in data sets, kinematic cuts, and analysis frameworks, 
it is still worth highlighting key features of our current understanding regarding 
individual parton-to-hadron FFs.
As can be observed in Fig.~\ref{fig:FFs_mixed}, a clear pattern emerges in both the uncertainty
estimates and the general agreement among the results from different collaborations.
Unsurprisingly, the best-determined distribution corresponds to the flavor singlet 
combination of FFs in Eq.~\ref{eq:singlet}, which is primarily probed via very precise SIA data
at $\mu=M_Z$. This combination not only features the smallest uncertainties but 
also exhibits remarkable agreement among the results from different fits.
The next best constrained combinations of FFs correspond to $u+\bar u$ and,
to a somewhat lesser extent, $d+\bar d$ for $z\gtrsim 0.2$, 
where SIDIS measurements are available and provide some handle on the flavor separation. 
For smaller values of the momentum fraction $z$, these distributions are not independently probed by data, 
and the result of the fit corresponds, to some degree, to an extrapolation of the behavior at larger $z$ behavior, 
heavily dependent on the adopted parametrization. 
In this region of $z$, the differences between global fits become in general more sizable, and are 
not always covered by the reported uncertainty bands, indicating some potential shortcomings 
in their estimation due to a lack of flexibility in the functional form or an insufficient sampling of the 
vast parameter space. 
The peculiar behavior for $D_{d+ \bar d}^{\pi^+}$ in the case of MAPFF1.0 can be understood from the
assumptions made about isospin symmetry, see Sec.~\ref{subsec:symmetries}. 
While BDSSV22, NPC23, and JAM20-SIDIS largely impose it up to
a small breaking, implying that $D_{u+\bar u}^{\pi^+}\approx D_{d+\bar d}^{\pi^+}$, both combinations
are parametrized independently by MAPFF1.0 and, hence, are solely constrained by data. Due to its
smaller electric charge, the total down-quark contribution has a smaller weight in SIDIS data taken
on a proton target as compared to the total up-quark contribution, which leads to larger uncertainties.
Similar comments can be made regarding the unfavored distributions $D_{s+ \bar s}^{\pi^+}$ 
and, in particular, $D_{\bar u}^{\pi^+}$ shown in the lower panels of Fig.~\ref{fig:FFs_mixed}.
While there is still some overlap of the different fits within their uncertainty bands in some regions
of $z$, the general agreement is much poorer than for the results shown in the upper row, 
especially for $D_{\bar u}^{\pi^+}$. Clearly, the lack of viable constraints from data 
translates into larger uncertainties and a strong dependence on the assumptions made in the fit

The largest differences between different fits can be observed for the gluon FF though.
This is not unexpected considering none of the extractions shown in Fig.~\ref{fig:FFs_mixed} includes data on 
single-inclusive hadron production in $pp$ collisions that is most sensitive to $D_{g}^{\pi^+}$. 
This leaves $D_{g}^{\pi^+}$ to be probed solely by scaling violations and sub-leading 
contributions to SIA and SIDIS.
Again, the differences between the fits fall outside the quoted uncertainty bands.
It should be noted, though, that large differences between the gluon FFs persist even when comparing 
NLO global analyses that include $pp$ data. It turns out that the gluon FF is extremely sensitive to the 
specific selection of data sets included in the fit. 
In particular, large tensions between measurements taken at RHIC and the LHC at much higher c.m.s.~energies
up to $\sqrt{S}=13\,\mathrm{TeV}$ are known to exist, with a good description of either data set crucially 
depending on the behavior of the fitted gluon FF \cite{dEnterria:2013sgr, Arleo:2010kw}. 
These tensions were to some degree mitigated in analyses simultaneously fitting 
both RHIC and LHC data by either employing a larger cut on the minimum allowed $p_T$ of 
the hadron \cite{deFlorian:2014xna}, exploiting the arbitrariness in the choice of
the factorization scale $\mu_f$ \cite{Borsa:2021ran, Borsa:2023zxk} or by 
including the sizable scale dependence as an additional source of uncertainty \cite{Gao:2024dbv}.
Since the scale dependence for hadron production in $pp$ collisions at NLO accuracy remains sizable,
in particular, at not too large values of $p_T$, including the now available NNLO QCD corrections
\cite{Czakon:2025yti} into the global analyses framework is the next and most critical step forward
in improving our knowledge of FFs. By the time of writing, no such analysis yet exists,


%
%

%

%

\section{Summary and Outlook}\label{chap5:summary}
We have presented an up-to-date account of the hadronization of quarks and gluons in QCD
through a set of non-perturbative but universal FFs. The underlying framework is
deeply rooted in nontrivial factorization theorems that enable us to compute a wealth of
experimentally measurable cross sections for identified hadrons systematically as a perturbative series.
The nonexistence of free quarks and gluons and, hence, the formation of colorless hadrons, is a fundamental 
and inevitable consequence of the long-distance properties of the strong interactions described by QCD.
While FFs do not provide a deeper understanding of confinement they 
conveniently parametrize hadronization in way that allows us to make quantitative
predictions in pQCD that can be verified and improved by experiment.

Compared to previous reviews of FFs \cite{Albino:2008gy,Metz:2016swz} considerable
progress has been made both in theory and in experiment. 
By now, most of the necessary theoretical ingredients for the scale evolution
of FFs and the evaluation of cross sections sensitive to hadronization are
available up to NNLO accurary, often accompanied by additional all-order
resummations up to NNLL or sometimes even beyond. Clearly, the framework
to study hadronization in terms of universal collinear parton-to-hadron FFs 
has matured to a level that permits detailed quantitative analyses of FFs 
including estimates of their uncertainties and critical inspections of the 
applicability and limitations of the foundational notion of factorization.
This is accompanied by a plethora of new experimental results mainly, but not
exclusively, from the LHC. Ever improving detector capabilities to
measure and identify different hadron species or even hadrons inside jets
have pushed the theory community to provide the necessary formalism 
to analyze all these all measurements using more and more refined tools.

Several different groups are actively pursuing the determination of
FFs through global QCD analyses based on different and always improving
and refined methodologies, which is crucial to arrive at faithful results with
trustworthy uncertainties in the end. We expect that future determinations of FFs
will more routinely exploit less inclusive probes such as in-jet or associated 
hadron production.
In short order, NNLO accuracy for all ingredients will be the standard
in fits of FFs, perhaps supplemented by occasional all-order resummations
if needed. This appears to be sufficient framework to account for 
future data from the LHC and eagerly awaited first results from 
the Electron-Ion Collider which should come online towards the end of 
the next decade.

Finally, there are many important avenues related to the hadronization of
quarks and gluons that could not be pursued in this short review,
most importantly the rich phenomenology and theory behind the transverse momentum 
dependence of FFs but also the production of heavier, charm and bottom flavored mesons 
and the physics behind hadronization models used in 
current event generators which are at the heart of all measurements.

\begin{ack}[Acknowledgments]

We thank Werner Vogelsang for reading this manuscript and for useful discussions. This work has been funded by the Deutsche Forschungsgemeinschaft (DFG, German Research Foundation)
through the Research Unit FOR 2926 (Project No. 409651613).

\end{ack}

\bibliographystyle{Harvard}
\bibliography{reference}

\end{document}